\documentclass[prc,showpacs,showkeys,preprint]{revtex4}
\usepackage{epsfig,graphicx,float,color}
\usepackage{amssymb}
\usepackage[normalem]{ulem}

\begin{document}

\title{Excited-state quantum phase transitions in a two-fluid Lipkin model}
\author{J.E.~Garc\'{\i}a-Ramos$^1$, P.~P\'erez-Fern\'andez$^2$, J.M~Arias$^3$}  
\affiliation{
  $^1$Departamento de  Ciencias Integradas, Universidad de Huelva,
  21071 Huelva, Spain and 
  Instituto Carlos I de F\'{\i}sica Te\'orica y Computacional,
  Universidad de Granada, Fuentenueva s/n, 18071 Granada, Spain\\
  $^{2}$Departamento de  F\'{\i}sica Aplicada III, Escuela T\'ecnica
  Superior de Ingenier\'{\i}a, Universidad de Sevilla, Sevilla,
  Spain\\ 
  $^{3}$Departamento de F\'{\i}sica At\'omica, Molecular y Nuclear,
  Facultad de F\'{\i}sica, Universidad de Sevilla, Apartado~1065,
  41080 Sevilla, Spain} 
\begin{abstract} 
\begin{description}
\item [Background:] Composed systems have became of great interest in
  the framework of the ground state quantum phase transitions (QPTs) and
  many of their properties have been studied in detail. However, in
  these systems the study of the so called excited-state quantum phase
  transitions (ESQPTs) have not received so much attention.
\item [Purpose:] A quantum analysis of the ESQPTs 
  in the two-fluid Lipkin model is presented in this
  work. The study is performed through the Hamiltonian diagonalization
  for selected values of the control parameters in order to cover the
  most interesting regions of the system phase diagram.
\item [Method:] A Hamiltonian that resembles the consistent-Q
  Hamiltonian of the interacting boson model (IBM) is diagonalized for
  selected values of the parameters and properties such as the density
  of states, the Peres lattices, the nearest-neighbor spacing
  distribution, and the participation ratio are analyzed.
\item [Results:] An overview of the spectrum of the two-fluid Lipkin
  model for selected positions in the phase diagram has been obtained.
  The location of the excited-state quantum phase transition can be
  easily singled out with the Peres lattice, with the
  nearest-neighbor spacing distribution, with Poincar\'e
    sections or with the participation ratio.
\item [Conclusions:] This study completes the analysis of QPTs for the
  two-fluid Lipkin model, extending the previous study to excited
  states. The ESQPT signatures in composed systems behave in the same
  way as in single ones, although the evidences of their presence can
  be sometimes blurred. The Peres lattice turns out to be a convenient
  tool to look into the position of the ESQPT and to define the concept
  of phase in the excited states realm.
\end{description}
\end{abstract}

\pacs{21.60.Fw, 02.30.Oz, 05.70.Fh, 64.60.F-}

\keywords{Lipkin model, two-fluid system, excited-state quantum phase transitions}
\maketitle

\section{Introduction}
\label{sec-intro}
During almost twenty years, quantum phase transitions (QPTs) have been a
hot topic in different areas of quantum many-body physics. On one
hand, QPTs in Nuclear Physics have been deeply studied
\cite{Cast07,Cejn09,Cejn10}, from both theoretical and experimental points
of view. On the other hand, other fields such as Molecular
Physics \cite{Iach08,Pere11a}, Quantum Optics \cite{Emar,Amic08} or
Solid State Physics \cite{Sach11} put forward the QPTs studies.

The well known thermodynamic phase transitions develop in systems with
an infinite number of particles, {\it i.e.}, in the thermodynamic
limit, in this sense they are called classical phase transitions. QPTs
are phenomena similar to classical phase transitions but differ in
that QPTs take place at zero temperature. In a broad sense, QPTs
appear in Hamiltonians that can be split into two parts, each of them
presenting a different symmetry. In this situation a simple
transitional Hamiltonian can be written as a function of one control
parameter that governs the change in the system from one symmetry to
the other,
\begin{equation}
H(\xi)=\xi\cdot H(\mbox{symmetry}_1)+ (1-\xi)\cdot
H(\mbox{symmetry}_2).
\label{Hschem}
\end{equation}
The phase of the system is characterized by a parameter, usually
called order parameter, that is zero in one phase and different from
zero in the other. A QPT is characterized by a sudden change in the
value of the order parameter for a small variation around a particular
value, $\xi_c,$ of the control parameter, $\xi$. The value $\xi_c$
where the QPT develops is known as the critical point and marks when
the system undergoes a structural change from symmetry$_1$ to
symmetry$_2$.

An appealing step forward in the QPT concept is its extension to
composed systems, {\it i.e.}, systems with different species of
components. The simplest case is a quantum system with two of such
species or fluids. One interesting case is the composed boson-fermion
system \cite{Alon05,Alon07a,Alon07b}, although here we will focus in a
two fluid model in which the two species are bosons and are
represented by creation and annihilation boson operators that fulfill
the usual boson commutation relations. In this framework, it is worth
to mention the very first studies in Nuclear Physics of two-fluid
systems \cite{Aria04,Capr04,Capri05}, conducted for the proton-neutron
interacting boson model. In similar schemes, the bending dynamics of
tetratomic molecules has also been studied with a two-fluid bosonic
model where each fluid is associated with a bender
\cite{Pere12,Lare14}. Other two-fluid systems, that still today act as
landmarks, are the Dicke \cite{Dick54} and the Jaynes-Cumming
\cite{Jayn63} models for which the two fluids correspond to photons
and atoms. Finally, another simple composed model is the two-fluid
Lipkin model \cite{Lipk65}, where the fluids may correspond to two
species of atoms or to two vibrational modes. Dicke and Jaynes-Cumming
models are algebraically connected with the two-fluid Lipkin
model. Indeed, the dynamical algebra of the double Lipkin model is
$u(2)\otimes u(2)$ and one can go to Dicke and Jaynes-Cumming models
through a {\it contraction} from the $u(2)$ Lie algebra to the
Heisenberg-Weyl one $hw(1)$ \cite{Iach06}.

The aim of this work is to extend the study of the two-fluid
Lipkin model, whose QPTs and phase diagram were studied in detail in
\cite{Garc16}, to the excited states realm, in other words, to study the
excited-state quantum phase transitions (ESQPTs) of the model. The
term ESQPT was first coined in \cite{Cejn06} and studied in detail in
\cite{Capr08}. In \cite{Capr08} it is stated that ``The infinite level
density, moreover, propagates to higher excitation energy ... hence
the concept of a continuation of the QPT to excited states''. Consequently, an ESQPT
can be understood as the propagation of the QPT to excited
states. Moreover, an ESQPT is deeply connected to the existence of a
barrier in the potential energy surface and, in particular, with the
height of the barrier. The presence of an ESQPT is marked by the
presence of a singularity in the density of states, but it is also
known to affect the structure of the states. In fact, these change,
from being non-symmetric or deformed to symmetric or spherical (or
viceversa), when crossing the ESQPT. Another relevant fact of the ESQPT
is that it seems to be related to a change from a regular to a chaotic regime,
although this point is still an open question.

In this manuscript, the study of the phase diagram for the two-fluid
Lipkin model is completed by extending the work of Ref.\ \cite{Garc16} to the excited
states.  The paper is organized as follows: in Section \ref{sec-model}
the algebraic structure and the model Hamiltonian are outlined. In
Section \ref{sec-ph-diagram} the main ingredients of its  
phase diagram are revised.  In Section \ref{sec-esqpt}
the ESQPT concept is introduced and, in particular, the
main tools used to study the onset of an ESQPT are commented. In Section \ref{sec-cases} 
the main outcome of this work is presented and we study in detail the
onset of ESQPTs in some particular cases. Finally, Section
\ref{sec-conclusions} stands for the summary and conclusions. 

\section{The Lipkin model and its two-fluid extension}
\label{sec-model}

The Lipkin model was proposed in the 1960's by Lipkin, Meshkov, and Glick
\cite{Lipk65} as a simple exactly solvable model to check the
validity and limitations of different approximation methods used in Nuclear Physics. Since
then, the model has been applied to other fields and many examples of
its use can be found in the literature.

\subsection{Algebraic structure}
\label{sec-algebra}

Using a boson representation, the Lipkin model is built in terms of
scalar bosons that can occupy two non-degenerated energy levels
labeled by $s$ and $t$. In the case of a single fluid, the building
blocks are the creation $s^\dag$, $t^\dag$, and annihilation $s$, $t$,
boson operators. The four possible bilinear products of one creation
and one annihilation boson operator generate the $u(2)$ algebra. The
next step to obtain the two-fluid Lipkin model is to combine two
coupled Lipkin structures. In this model, there are two boson families
identified by a subindex, $s_1^\dag$, $t_1^\dag$ and $s_2^\dag$,
$t_2^\dag$, and the corresponding dynamical algebra will be
$u_1(2)\otimes u_2(2)$, whose generators are: $s^\dag_i s_i$,
$s^\dag_i t_i$, $t^\dag_i s_i$, and $t^\dag_i t_i$, for $i=1,2$.

A detailed description of the $u_1(2) \otimes u_2(2)$ algebraic
structure can be found in \cite{Fran94}. Here, we simply summarize
some features that will be of interest along this work. Starting from the dynamical algebra $u_1(2) \otimes
u_2(2)$, the possible subalgebras chains are four. However, for us only
two of them, in which there is an early coupling of the
dynamical algebras into the direct-sum subalgebra $u_{12}(2)$ (or $su_{12}(2)$), are relevant, 
\begin{equation}
\begin{array}{cccccc}
u_1(2)\otimes u_2(2)&\supset&u_{12}(2)&\supset& u_{12}(1) & \\
\downarrow& &\downarrow& &\downarrow\\
N_1 \otimes N_2& &[h,h']& &n_t & \rightarrow \textrm{basis} ~~ |N_1~  N_2~ h~ n_t \rangle \\
\end{array},
\end{equation}
where the labels of the irreps verify the following branching rules: 
$h+h'=N_1+N_2$, $h \ge h'$, $1/2(N_1+N_2+h'-h)\le n_t\le
1/2(N_1+N_2+h'-h)$,  and
\begin{equation}
\begin{array}{cccccccc}
u_1(2)\otimes u_2(2)&\supset&su_1(2)\otimes su_2(2)&\supset&su_{12}(2)&\supset& so_{12}(2)& \\
\downarrow& &\downarrow& &\downarrow& &\downarrow\\
N_1 \otimes N_2& &j_1 \otimes j_2& &j& &\mu & \rightarrow  \textrm{basis}~~ |j_1~  j_2~ j~ \mu \rangle \\
\end{array},
\end{equation}
where $j_i=N_i/2$,  $j=1/2(N_1+N_2),1/2(N_1+N_2)-1, ...,
1/2|N_1-N_2|$, $-j\le \mu\le j$, and $j=1/2(h-h')$. $|...\rangle$ stands for
the basis state in the corresponding dynamical symmetry.

The Hamiltonian used in this work is inspired in the
consistent-Q-formalism of the interacting boson model \cite{Cast88}. 
The Hamiltonian can be written as,
\begin{equation}
H= x \left(n_{t_1}+n_{t_2} \right)-\frac{1-x}{N_1+N_2}
Q^{(y_1,y_2 )}\cdot Q^{(y_1,y_2 )}    
\label{Hcqf}
\end{equation} 
where
\begin{eqnarray}
\label{ni} n_{t_i}&=&t_{i }^{\dagger }t_{i },\\ 
Q^{(y_1,y_2)}&=&\left(Q_1^{y_1}+
Q_2^{y_2}\right), \\ 
\label{Qi} Q_i^{y_i}&=& 
s_{i }^{\dagger } t_{i } + t_{i }^{\dagger }s_{i }+ 
y_i \left( t_{i}^{\dagger }{t}_{i}\right).
\end{eqnarray}
The Hamiltonian (\ref{Hcqf}) has three control parameters ($x, y_1$,
and $y_2$). The model has two order parameters $\beta_1$ and $\beta_2$
associated to each fluid: $\beta_i=0$ values indicate the symmetric or
spherical phase for fluid $i$, while values different from zero
characterize non-symmetric or deformed phases. Due to the behaviour of
the bosons under parity, the Hamiltonian (\ref{Hcqf}) is, in general,
non-parity conserving, except for $y_1=y_2=0$.  

The Hamiltonian (\ref{Hcqf}) is a mixture of dynamical symmetries of the problem,
particularly $u_{12}(1)$ for $x=1$, and $so_{12}(2)$ for $x=0$ and $y_1=y_2=0$. 
This form is specially suitable to study QPTs, because one can
associate a \textit{symmetric} (spherical) phase to the first term of
the Hamiltonian and a \textit{non-symmetric} (deformed) shape to the
second term. Moreover, depending on the values of $y_1$ and $y_2$,
different kinds of deformation are generated. For us, it is specially
important the case $y_1=y_2$, for which the dynamical algebra of the
Hamiltonian will be $u_{12}(2)$ ($su_{12}(2)$) and the states will
belong to a single $[h,h']$ Young tableau (with a well defined value of
$j$). In this case, one can separate the spectrum in
families with given $j$-values. Note that although $j$ has the
properties of an angular momentum, it is not an orbital angular
momentum in the sense of $L$ in the interacting boson model or in the
vibron model. It is, indeed, similar to the concept of F-spin in the
proton-neutron interacting boson model \cite{Iach87}.

\subsection{The phase diagram}
\label{sec-ph-diagram}
\begin{figure}[hbt]
  \centering
\includegraphics[width=0.70\linewidth]{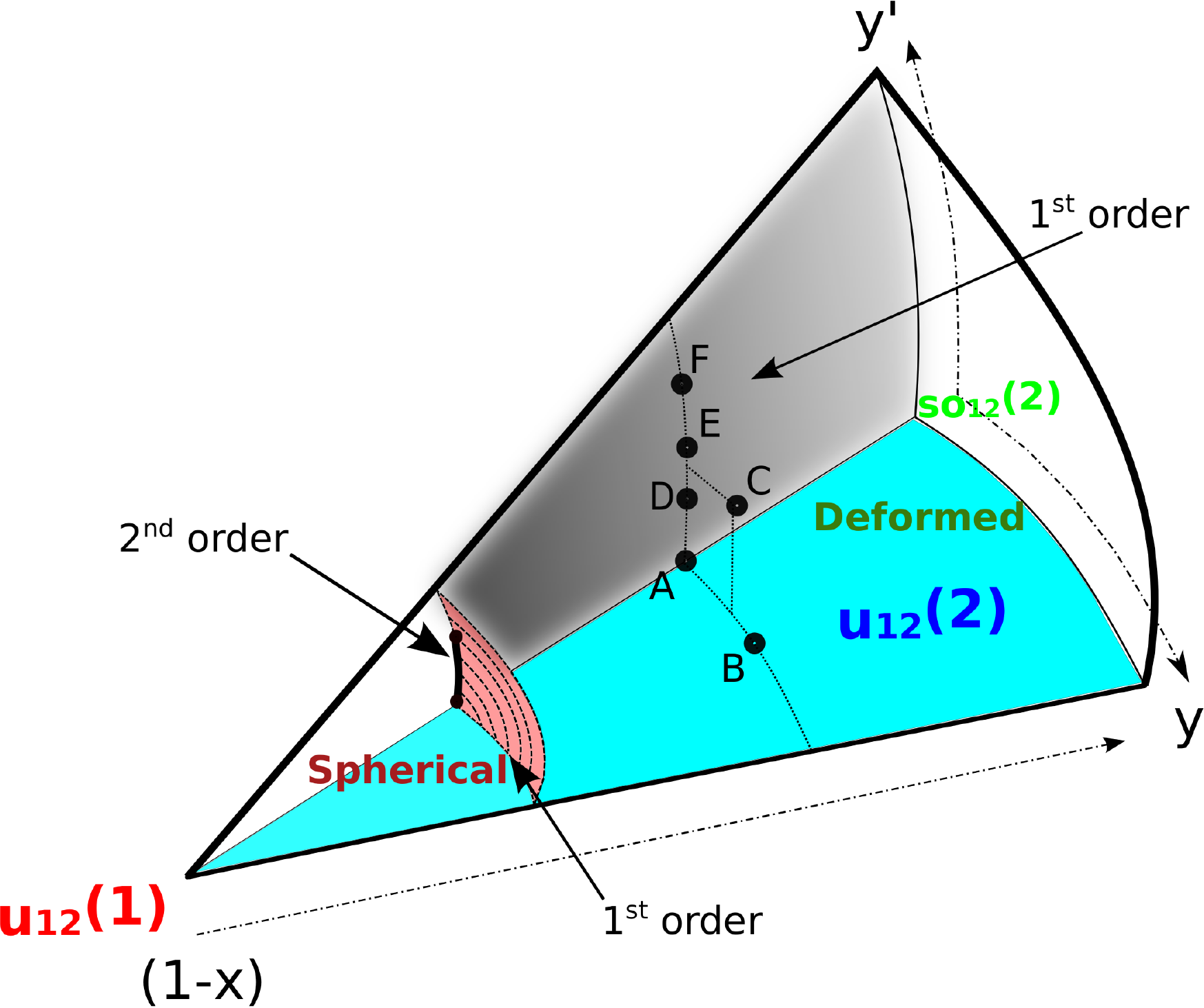}
\caption{Phase diagram of the consistent-Q like
  two-fluid Lipkin model. In the diagram the different phases are
  represented: spherical and deformed, the first order QPT surfaces
  and the second order QPT line. Moreover the relevant control
  parameters ($x$, $y=(y_1+y_2)/2$, and $y^\prime=(y_1-y_2)/2$) and
  dynamical symmetries also are shown. The marked points correspond to
  the cases studied in Section \ref{sec-cases} (see text).} 
\label{fig-dia}
\end{figure}

The combination of numerical calculations with analytical results, as
shown in \cite{Garc16}, provides the phase diagram depicted in
Fig.~\ref{fig-dia}. In that reference an essential order parameter
$\beta=(\beta_1+\beta_2)/2$ is defined and it is the parameter that
characterizes the different phases: spherical means $\beta=0$, while
two different deformations can appear, $\beta>0$ and $\beta<0$. Note
that the order parameter $\beta$ is equivalent to $\beta_a$ as it
appears in Ref.\cite{Garc16}.

The Hamiltonian (\ref{Hcqf}) can be more conveniently rewritten in terms of control
parameters $y=(y_1+y_2)/2$, and $y^\prime=(y_1-y_2)/2$. In Fig.~\ref{fig-dia} the
phase diagram of the model in the space of coordinates $x$, $y$, and $y'$ is presented.
There, a first order phase transition
surface (in red) separates the symmetric (spherical, $\beta=0$) and
non-symmetric (deformed) phases (the one shown here corresponds to
$\beta>0$). This first order phase transition region appears in many
other models (quantum cusp, interacting boson model, Dicke and
Jaynes-Cumming, among others) and essentially stems from the
competition between single particle terms that lead to spherical
shapes and two-body interactions that lead to deformed
configurations. Therefore, this phase transition is related with the
evolution and competition of the spherical and the deformed minima
(there is a region of coexistence of minima and the critical point is
defined by the degeneration of both minima). This surface contains a
line (black thick line in the figure) that runs from $y=y'=0$ to
$y=0$, $y'=1$ and corresponds to a second order phase transition. This
line is, in fact, a triple point where three degenerated minima
coexist (one spherical, and two deformed with different
deformation). The point $y=0$, $y'=1$ shows a unique behaviour because
presents a divergence \cite{Garc16} in the second derivative of the
energy with respect to the control parameter. At this point, the
spinodal and the antispinodal lines merge with the first order phase
transition surface giving rise to a tricritical point.

The phase diagram shown in Fig.~\ref{fig-dia} can be completed by
extending it to negative values of $y$ and $y'$. In particular the
negative $y$ values imply deformation $\beta<0$. The vertical surface
(in grey), separates deformed phases with different signs in the value
of the order parameter $\beta$. This is a first order phase transition
surface. This phase transition is connected with the existence of two
deformed minima, $\beta<0$ and $\beta>0$, that eventually can become
degenerated (the grey surface in the figure). Note that in this
situation, the potential is symmetric under an appropriate interchange
of the two shape variables ($\beta_1$ and $\beta_2$, order
parameters). At each side of the surface two deformed minima coexist
but in the side shown in Fig.~\ref{fig-dia} the absolute minimum
corresponds to
$\beta>0$ while at the other side the absolute minimum appears for  $\beta<0$.
This situation is different to what happens at the phase transition
surface separating spherical ($\beta=0$) and deformed ($\beta>0$)
phases. In this case, the system transits from a single spherical
minimum, $\beta=0$, to a deformed minimum ($\beta>0$) through a
coexistence region. In the interacting boson model the first situation
corresponds to the QPT between the $SU(3)$ and the $\overline{SU(3)}$
limits \cite{Joli01}, while the second corresponds to the QPT
appearing when passing from the $U(5)$ limit to the $SU(3)-O(6)$ line
\cite{Iach87}.

\section{Excited-state quantum phase transitions}
\label{sec-esqpt}
In many-body quantum systems the presence of a ground state QPT can
give rise to an ESQPT when using the excitation energy as a control
parameter \cite{Capr08}. Once in the deformed phase, keeping the
control parameters fixed, one can go up in energy and look into a
magnitude that marks the presence of a quantum phase transition in the
excited states. This magnitude can be the density of states which is
expected to have some kind of singularity when reaching the energy at
which an ESQPT develops.  In particular, this is the case for the
one-fluid Lipkin model, where the QPT for the ground state seems to
``propagate'' to the excited states.  An ESQPT associated to second
order QPT is defined as a singularity in the density of states or in
one of its derivatives. The kind of singularity depends on the number
of degrees of freedom of the system in the semiclassical limit
\cite{Stra14,Stra15}. In particular, the Lipkin model, which has a
single degree of freedom, presents a $\lambda$ singularity in the
density of states at the excitation energy corresponding to the
ESQPT. The Dicke model, with two degrees of freedom, presents at the
ESQPT critical energy a discontinuity in the first derivative of the
density of states \cite{Stra14}. The sudden increase in the density of
states is related with the presence of a maximum in the potential
energy surface of the system. Starting from the bottom of the
potential the states bunch up when reaching the maximum of the
potential giving rise to an increase of the density of states.  On the
other hand, when the QPT is of first order, besides the previous
behaviour, linked to the existence of a maximum, the presence of a new
family of states related with the existence of an extra local minimum
induces a finite increase in the density of states and, therefore, a
discontinuity in the density of states. The energy at which the second
family of states appears in the spectrum corresponds to the energy of
the second minimum.

\begin{figure}[hbt]
\includegraphics[width=0.90\linewidth]{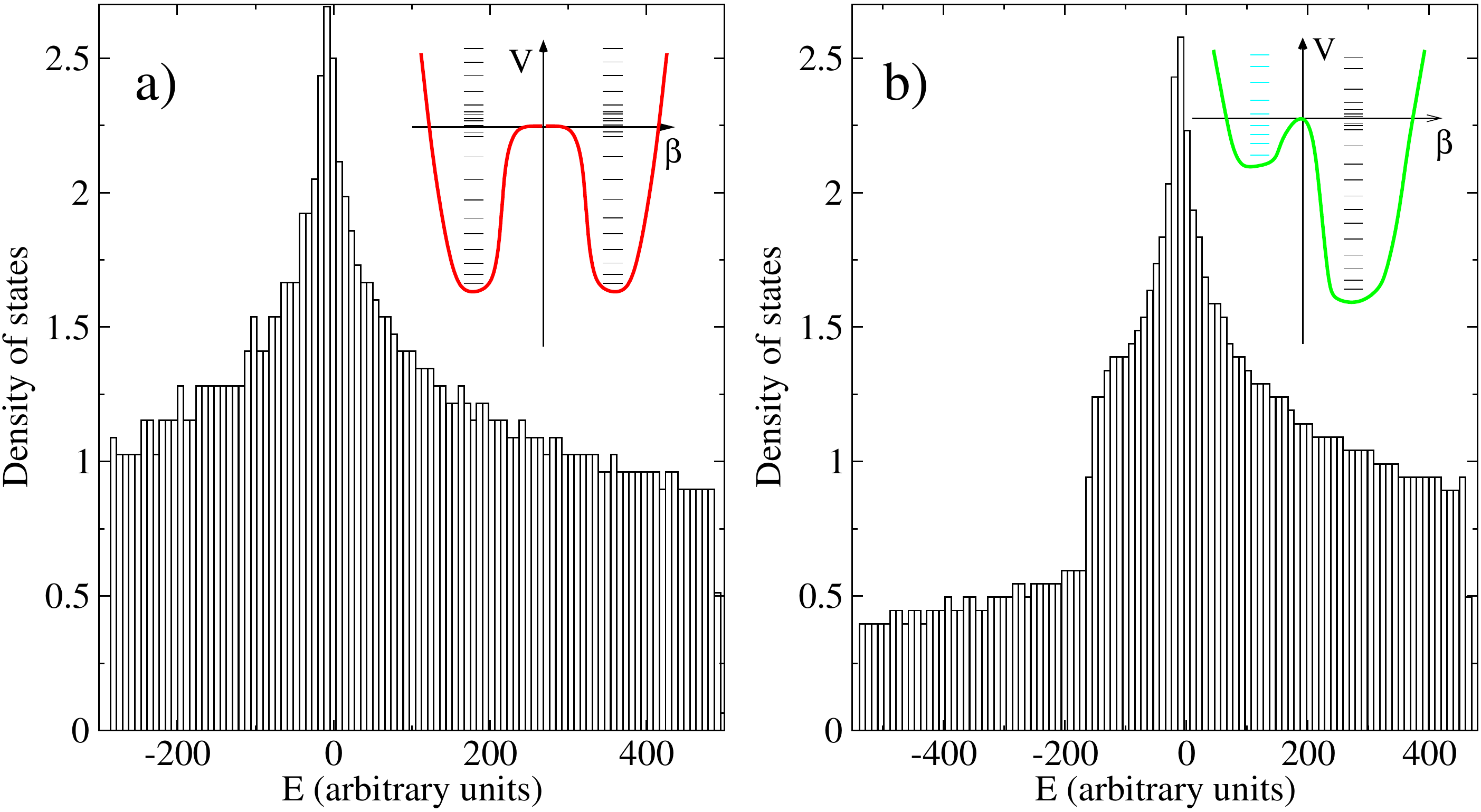}
\caption{Density of states for a one-fluid Lipkin Hamiltonian $H= x
  n_{t}-\frac{1-x}{N} Q^{(y)} \cdot Q^{(y)}$ ($n_t$ and $Q^{(y)}$
  defined in (\ref{ni}) and (\ref{Qi}) with $x=1/2$, $y=0$, and
  $N=1000$ (panel (a)) and with $x=1/2$, $y=1/2$, and $N=1000$ (panel
  (b)); the insets show schematically the corresponding  potential
  energy surface in both cases. }    
\label{fig-qualitative-ESQPT}
\end{figure}

To explain more in detail the connection between the presence of a
maximum in the potential energy surface and the onset of an ESQPT, we
present in Fig.~\ref{fig-qualitative-ESQPT} two calculations for large
number of bosons, $N=1000$, evaluating the density of states for the
single Lipkin Hamiltonian. In panel (a) we perform a calculation for a
Hamiltonian in the deformed phase. The Hamiltonian represents a
situation in which the ground state is deformed but comes from the
evolution of a spherical ground state through a second order phase
transition, this means that there is a maximum at zero deformation
(see inset of Fig.~\ref{fig-qualitative-ESQPT}a). Note that there are,
in fact, two degenerated minima that give rise to degenerated doublets
below the ESQPT. The existence of the maximum induces many states
bunching together in the spectrum, leading, therefore, to a sudden
increase (a singularity in the thermodynamic limit) in the density of
states at the energy of the maximum (this is seen at zero energy in
panel (a)). In order to show the difference with the case of a first
order phase transition, in panel (b) a calculation of the density of
levels is presented for a Hamiltonian owning a deformed ground state
that now comes from the evolution of a spherical ground state through
a first order phase transition. This means that at the QPT there is
coexistence of spherical and deformed minima. This fact has as a
consequence that far from the QPT, in the deformed phase, the system
presents a local deformed minimum too, being both deformed minima
separated by the spherical maximum (this can be seen in the inset of
Fig.~\ref{fig-qualitative-ESQPT}b). Starting from the bottom of the
potential and going up in energy, when the energy of the local minimum
is reached, a new family of states appears and it produces a finite
increase in the density of states (at energy around -200 in panel
(b)). Going upper in energy one reaches the maximum of the potential
and a new bunching up of the energy levels is observed in the
spectrum, producing a peak in the density of states (at zero energy in
panel (b)), this will give a lambda singularity in the thermodynamic
limit).
      
Let us now come to our case of interest, the two-fluid Lipkin model
given by Eq. (\ref{Hcqf}).  The Hamiltonian is such that for any set
of parameter values that lead to a deformed ground state, the energy
potential has always a maximum at $\beta=0$. The reason is the absence
of linear terms in the potential (see Eq.~(27) of \cite{Garc16}). This
also happens in the interacting boson model. Therefore, for such a
situation, it appears in the spectrum an ESQPT that is related with
the existence of a maximum in the potential energy surface at
$\beta=0$. As a consequence, for our Hamiltonian, the critical energy
of the ESQPT will be always zero because it corresponds to the value
of the potential energy surface at $\beta=0$.

Although the concept of phase is strictly defined for the ground
state, one can extend it to the excited states \cite{Capr08}, in the
sense that the excited states resemble the ground state with $\beta=0$
(symmetric states) or with $\beta\neq 0$ (non-symmetric states). As a
matter of fact, in Fig.~11 of Ref.~\cite{Capr08} one can clearly see
how the states can be divided into two families, each one with
reminiscences of one of the different Hamiltonian symmetries (see
Hamiltonian (\ref{Hschem})).  The states below the ESQPT energy can be
assigned to the non-symmetric (deformed) phase, while the ones above
the ESQPT energy are in the symmetric (spherical) phase, or viceversa.
 
\subsection{Some tools to study ESQPTs}

As we have already explained, an ESQPT has been mainly identified by some kind of
singularity in the density of states. However, this is neither the
only signal of the presence of an   
ESQPT, nor the most efficient, specially for the study of finite size systems. In this section,
a set of quantities that could serve as markers for ESQPT's are briefly discussed. 

\subsubsection{Density of states}

This is the most obvious tool to reveal the presence of an ESQPT,
however with this observable the presence of an ESQPT is not always so clear as
in Fig.~\ref{fig-qualitative-ESQPT}. For example, in the case of
systems with two degrees of freedom, the discontinuity  
appears in the derivative of the density of states and, therefore, it
is hard to be detected in a numerical calculation. The size of the
system is also an issue because it could hide the presence of an
ESQPT. For instance, a low number of states will hinder the detection of an ESQPT in any numerical 
calculation of density of states. Finally, the existence of an unnoticed symmetry in the 
Hamiltonian will change completely the observed behaviour in
Fig.~\ref{fig-qualitative-ESQPT} and the density of states will
correspond to the superposition of multiple belts with smoothed
singularities at an energy that is shifted, making much harder to
notice any singularity.

\subsubsection{Chaotic versus regular behaviour: Nearest-neighbor
  spacing  distribution, Poincar\'e sections, and  Peres lattices}

The interplay between ESQPTs and the onset of chaos has been studied
in \cite{Pere11b,Basta14,Chav16}.  It has been shown that,  in
% XXXXXXXXXXXXXXXXXXXXXX
particular cases, but not in general, the behaviour of
the nearest-neighbor spacing distribution (NNSD) suddenly changes from
regular to chaotic distribution when crossing the ESQPT energy. The
NNSD, thus, could be used as a signature of the presence of an ESQPT
at a given energy. However, it will become specially effective if the
character, either regular or chaotic, changes for all the states at
roughly the same energy, and there are no regular states coexisting
with the chaotic ones. The reason is that the NNSD provides a survey
that is not local but only valid within a certain range of energy.

For the Hamiltonian under study, the NNSD is not an ideal tool
% XXXXXXXXXXXXXXXXXXXXXX
because,  as it will be shown,
regular and chaotic states are not well separated in the
spectrum. However, as we will see, one can still get a clear signal
when crossing the energy of the ESQPT. To use the NNSD as a hint for
the presence of chaos we should remember that in integrable systems
the states follow a Poisson distribution $P_P(s)=exp(-s)$, while in
fully chaotic ones, they follow a Wigner distribution
$P_W(s)=\frac{\pi}{2} s \cdot exp(-\frac{\pi}{4}s^2)$, where
$s_i=(E_{i+1}-E_i)/\langle s \rangle$ is a normalized distance between
levels. We will measure the degree of chaos defining the quantity
\begin{equation} \eta=\frac{\sigma_s-\sigma_W}{\sigma_P-\sigma_W},
\label{eta}
\end{equation} where $\sigma_s=\langle s^2\rangle-\langle s\rangle^2$
is the variance of the analyzed spectrum, $\sigma_W=4/\pi-1$ is the
variance of the Wigner distribution, and $\sigma_P=1$ is the variance
of the Poisson distribution. Therefore, the system will be fully
regular for $\eta=1$, and fully chaotic for $\eta=0$. It is worth
noting that in order to calculate $\eta$ all the considered states
should have the same symmetry, {\it e.g.}, in the case of states with
given parity, all the states should have the same parity. To mark the
onset of chaos one can plot the value of $\eta$, calculated for a
certain number of states, as a function of the energy.

%XXXXXXXXXXXXXXXXXXXXXXXXXXXXXXXX
Another option to study the onset of chaos is to build the classical
counterpart of Hamiltonian (\ref{Hcqf}) and to study its semiclassical
dynamics \cite{Stra15}. To this end we define a coherent state as the one of
Ref.~\cite{Garc16}, but taking the deformation parameter, $\beta_k$,
as complex
and redefining  it, such that its absolute value is constrained to the
interval $[0,1]$,
$\tilde{\beta_k}\tilde{\beta_k}^*=\beta_k^2/(1+\beta_k^2)$, 
\begin{equation}
 |\tilde{\beta_1}, \tilde{\beta_2} \rangle =  \frac{1}{\sqrt{N_1!N_2!}} 
\left(\sqrt{1-\tilde{\beta_1}\tilde{\beta_1}^*}s_1^{\dagger}+\tilde{\beta_1}
t_1^{\dagger}\right)^{N_1}
\left(\sqrt{1-\tilde{\beta_2}\tilde{\beta_2}^*}s_2^{\dagger}+\tilde{\beta_2} 
t_2^{\dagger}\right)^{N_2}|0\rangle.
\end{equation}
The classical  limit of the system is then given by the expectation
value of the Hamiltonian: 
\begin{eqnarray}
H_{cl}(\tilde{\beta_1},\tilde{\beta_1}^*,\tilde{\beta_2},\tilde{\beta_2}^*)=\langle \tilde{\beta_1}, 
\tilde{\beta_2}|H|\tilde{\beta_1}, \tilde{\beta_2}\rangle .
\label{Hclas}
\end{eqnarray} 
From 
$\tilde{\beta_1}$ and $\tilde{\beta_2}$ one can define the canonical
variables $(q_1,p_1)$ and  $(q_2,p_2)$, such that they verify:
\begin{equation}
\tilde{\beta_k}=\frac{1}{\sqrt{2}}(q_k+\imath p_k),
\label{pq}
\end{equation} 
with $k=1,2$. Therefore, the classical Hamiltonian per particle,
$h_{cl}(q_1,p_1,q_2,p_2)=H_{cl}(q_1,p_1,q_2,p_2)/N$, for a system with
$N_1=N_2$ can be written as, 
\begin{eqnarray}
\nonumber
h_{cl}(q_1,p_1,q_2,p_2)&=&\frac{x}{4}(p_1^2+p_2^2+q_1^2+q_2^2)+\frac{x-1}{16} 
   \Bigg(2 q_1^2 \Big(p_1^2 (y_1^2-2)\\
\nonumber
     &+& 2 q_2 y_1
       \sqrt{-p_2^2-q_2^2+2}+p_2^2 y_1 y_2+q_2^2 y_1 y_2+4\Big) \\
\nonumber
     &+& 4 q_1  \sqrt{-p_1^2-q_1^2+2}\; \Big(p_1^2 y_1+ 2 q_2
     \sqrt{-p_2^2-q_2^2+2}\;+p_2^2 y_2+q_2^2 y_2\Big) \\
\nonumber
     &+& 2 q_2^2 \Big(p_1^2 y_1 y_2+p_2^2 (y_2^2-2)+4\Big)+4 q_2
   \sqrt{-p_2^2-q_2^2+2}\; (p_1^2 y_1+p_2^2
   y_2) \\
\nonumber
   &+&(p_1^2 y_1+p_2^2 y_2)^2+4
   q_1^3 y_1 \sqrt{-p_1^2-q_1^2+2}\\
   &+&4 q_2^3 y_2
   \sqrt{-p_2^2-q_2^2+2}\;+q_1^4
   (y_1^2-4)+q_2^4 (y_2^2-4)\Bigg),
\end{eqnarray} 
which is defined in a four-dimensional phase space, though  energy 
conservation allows to reduce it to three dimensions only.
The dynamics of the system is determined by Hamilton's equations:
\begin{eqnarray}
 \frac{dq_k}{dt}&=&\frac{\partial h_{cl}}{\partial p_k} \nonumber \\ 
 \frac{dp_k}{dt}&=&-\frac{\partial h_{cl}}{\partial q_k},
 \label{Hameq}
\end{eqnarray}
with $k=1,2$. Once Eqs.~(\ref{Hameq}) are solved numerically,
the nature of the motion, either regular or chaotic, can be easily 
depicted using the Poincar\'e sections, plotting the coordinate values of the
intersection in a given plane of different trayectories of
the system. In our case, we consider intersections with the plane
$p_2=0$ (note that $q_2$ is determined by energy conservation), and then
we collect the coordinates $q_1$-$p_1$.  
On the other hand, chaotic
regimes lead to trayectories that occupy the whole available phase space and
intersect randomly the Poincar\'e section. On the other hand, regular
situations correspond to trajectories that are constrained to toroidal
regions that generate closed curves when they intersect the Poincar\'e section.     
%XXXXXXXXXXXXXXXXXXXXXXXXXXXXXXXX

Finally, another very simple tool that is able to distinguish qualitatively
regular from chaotic states is the Peres lattice \cite{Pere84}.  A
Peres lattice provides a way to characterize states by simply
performing a diagram where each point corresponds to a single
state. In the diagram the matrix element of a convenient operator is
plotted versus the excitation energy.  In our case, an efficient
operator is $n_{t_i}$. Therefore, we will represent $\langle n_{t_1}
\rangle/N_1$ (or $\langle n_{t_2} \rangle/N_2$), whose values range
between $0$ and $1$, as a function of the excitation energy. In the
case of an energy region with a chaotic behaviour, the Peres lattice
provides a disordered distribution of points, while for regular
states, the pattern becomes ordered. Even in the case of coexistence
of regular and chaotic states in the same energy region, a Peres
lattice will allow to separate both families. Note that the Peres
lattice can also help to define the ``shape/phase'' of the excited
states, as explained in \cite{Capr08}. The characterization of the
excited states ``shape/phase'' is carried out by examination of the
pattern observed in the wave functions within a region, but not for a
single state. Peres lattices are an ideal tool to characterize
``shape/phase'' excited states within a given region.

\subsubsection{Participation ratio}  

A different way of studying the onset of an ESQPT is through the
analysis of the wave function. In particular, it is enlightening to
study how localized or delocalized is a given state since it has been
recently proven \cite{Sant15} that states nearby the critical energy
of an ESQPT are well localized, while the rest present sparse wave
functions.  A convenient quantity to study the structure of the wave
function is the participation ratio, $P$, which is defined, in a given
basis $\{|i \rangle \}$, for a wavefunction $|\psi_k\rangle=\sum_i
C^{(k)}_i |i\rangle$ as,
\begin{equation}
P^{(k)}=\frac{1}{dim}\frac{1}{\sum_i  |C^k_i|^4}, 
\end{equation} 
where $dim$ stands for the dimension of the Hilbert space.  This
function provides the degree of delocalization of a given state in a
particular basis, being, therefore, basis dependent.  For a well
localized state, $P$ is small ($1/dim$ in value), while it will become
large for a delocalized one ($1$ as maximum value).

In Ref.~\cite{Sant15}, the authors showed that using a $u(n)$ basis,
the participation ratio of the nearest eigenstate to the ESQPT shows a
marked dip that allows to localize very cleanly the ESQPT position. In
this work we will use the $u_1(1) \otimes u_2(1)$ basis in which the
quantum numbers $n_{t_1}$ and $n_{t_2}$ are specified, as explained in
Ref.~\cite{Garc16}.

\section{Cases of interest}
\label{sec-cases}

The first case we deal with is $y=0$, $y'=0$ ($y_1=0$,
$y_2=0$), this is the line from $u_{12}(1)$ to $so_{12}(2)$), with 
$x=1/2$, which corresponds to a deformed phase after the critical point
of a second order QPT at $x_c=4/5$. This case is represented with
point A in Fig.~\ref{fig-dia}. In the following, all the calculations will be performed
for $N_1=N_2=70$ bosons, which involves a dimension of $71\times
71=5041$. This Hamiltonian generates a deformed phase for the ground
state, {\it i.e.}, the order parameters are $\beta_1\neq 0$ and
$\beta_2\neq 0$. Because $y_1=y_2$ the system is symmetric under the
interchange of the index 1 and 2. Therefore, $u_{12}(2)$ will be the
dynamical algebra of the Hamiltonian and the states will belong to a
certain Young tableau, $[h,h']$, making possible to define an angular
momentum quantum number $j=1/2(h-h')$ with possible values
$j=1/2(N_1+N_2), 1/2(N_1+N_2)-1, ..., 1/2|N_1-N_2|$, as shown in
Sec.~\ref{sec-model}. As a consequence, the Hilbert space can be split
in subspaces with a given value of $j$ and $2j+1$ dimension.  This
fact should be taken into account for interpreting correctly the
coming figures, otherwise wrong conclusions can be reached.

\begin{figure}
\begin{tabular}{ccc}
\includegraphics[width=0.40\linewidth]{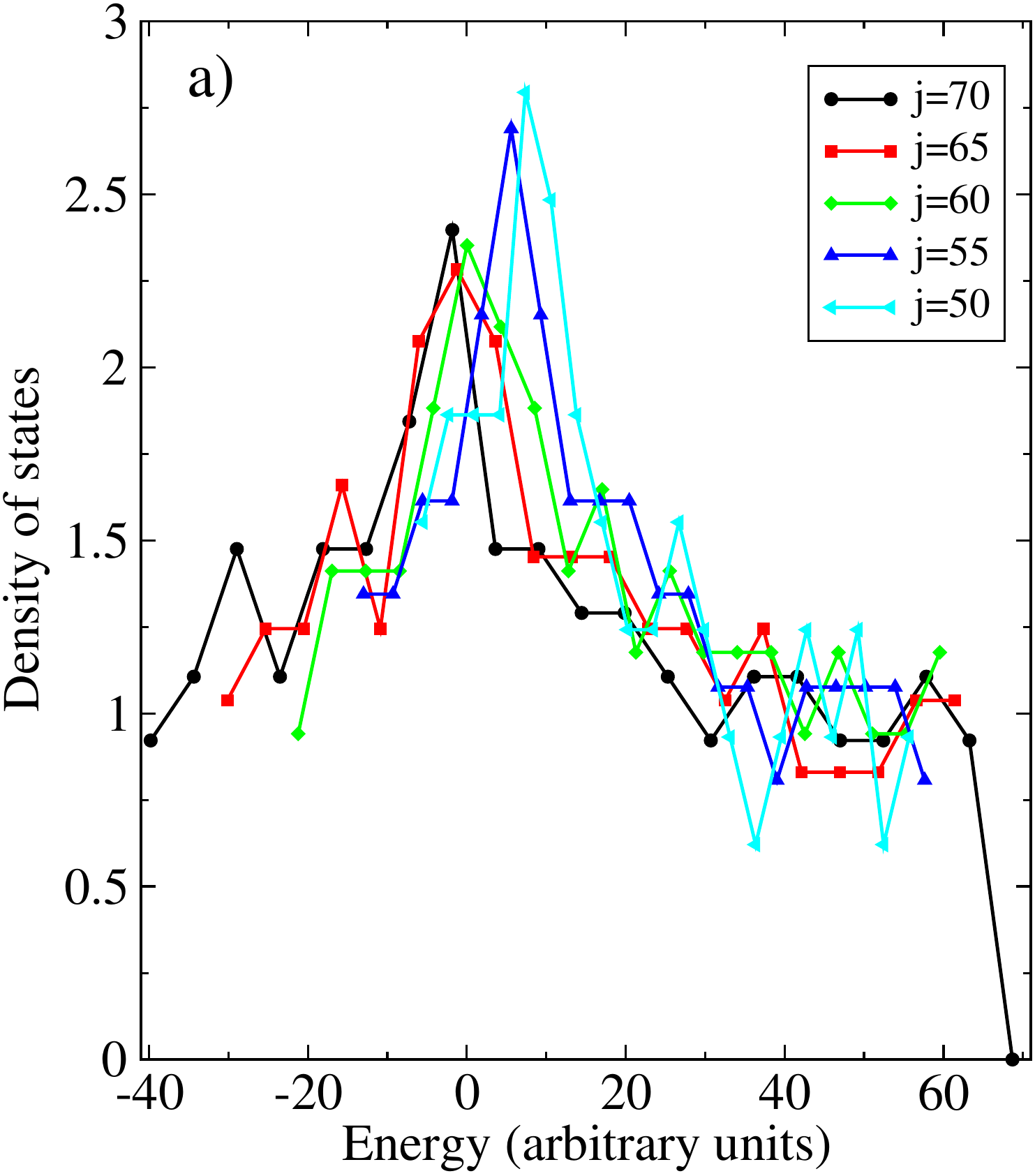}
&~~&
\includegraphics[width=0.40\linewidth]{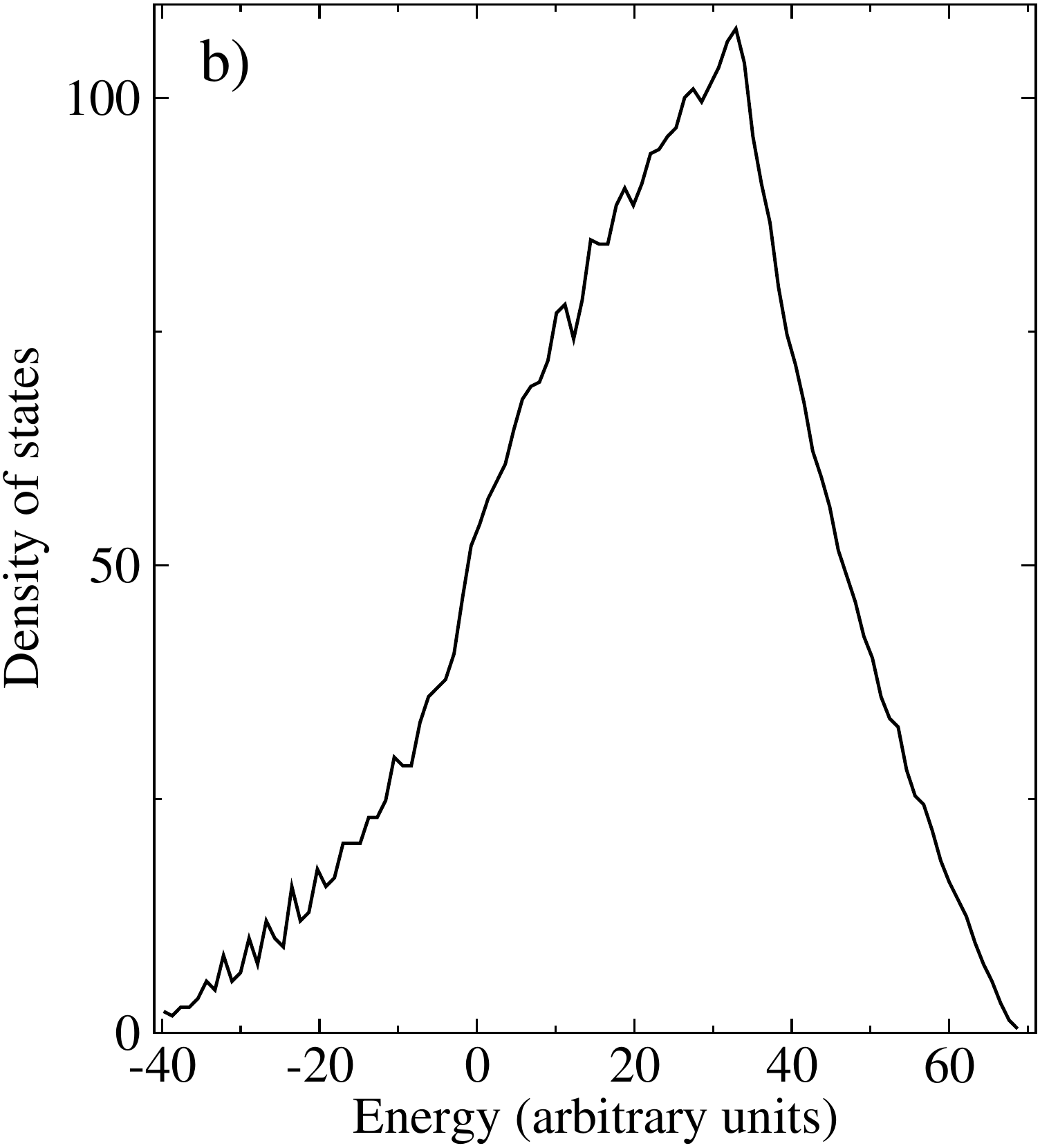}
\end{tabular}
\caption{Density of states as a function of the excitation energy for
  a Hamiltonian with parameters 
  $y=0$, $y'=0$, and $x=1/2$ and a number of bosons $N_1=N_2=70$. a)
  For selected values of $j$, b) for all the states without
  distinction in $j$.}   
\label{fig-density-16}
\end{figure}

Let us start with the study of the density of states
(Fig.~\ref{fig-density-16}), for which we already saw in
Fig.~\ref{fig-qualitative-ESQPT} how a lambda divergence is obtained
at the critical energy of the ESQPT. In Fig.~\ref{fig-density-16}a,
the value of the density of states for selected values of $j$ is
depicted (we use a different color for each $j$ value), for each of
them a peak is clearly observed, which is a 
precursor of a lambda divergence. Note also that as $j$ decreases the
energy of each peak is slightly shifted to higher values of the
energy. In panel (b), the density corresponds to the sum of the
individual densities presented in panel (a) and the consequence is
that the typical density values in panel (b) are much larger than in
panel (a) and that individual peaks no longer can be observed and,
indeed, the only observed peak is a broad one well above zero
energy. The energy of the peak at $E \approx 30$ is connected with the peak
position of the families with the smallest $j$ values, whose peaks are
located at the highest energy values. Consequently, for this case the
level density is not a good marker for the ESQPT.
 
\begin{figure}
\begin{tabular}{ccc}
\includegraphics[width=0.40\linewidth]{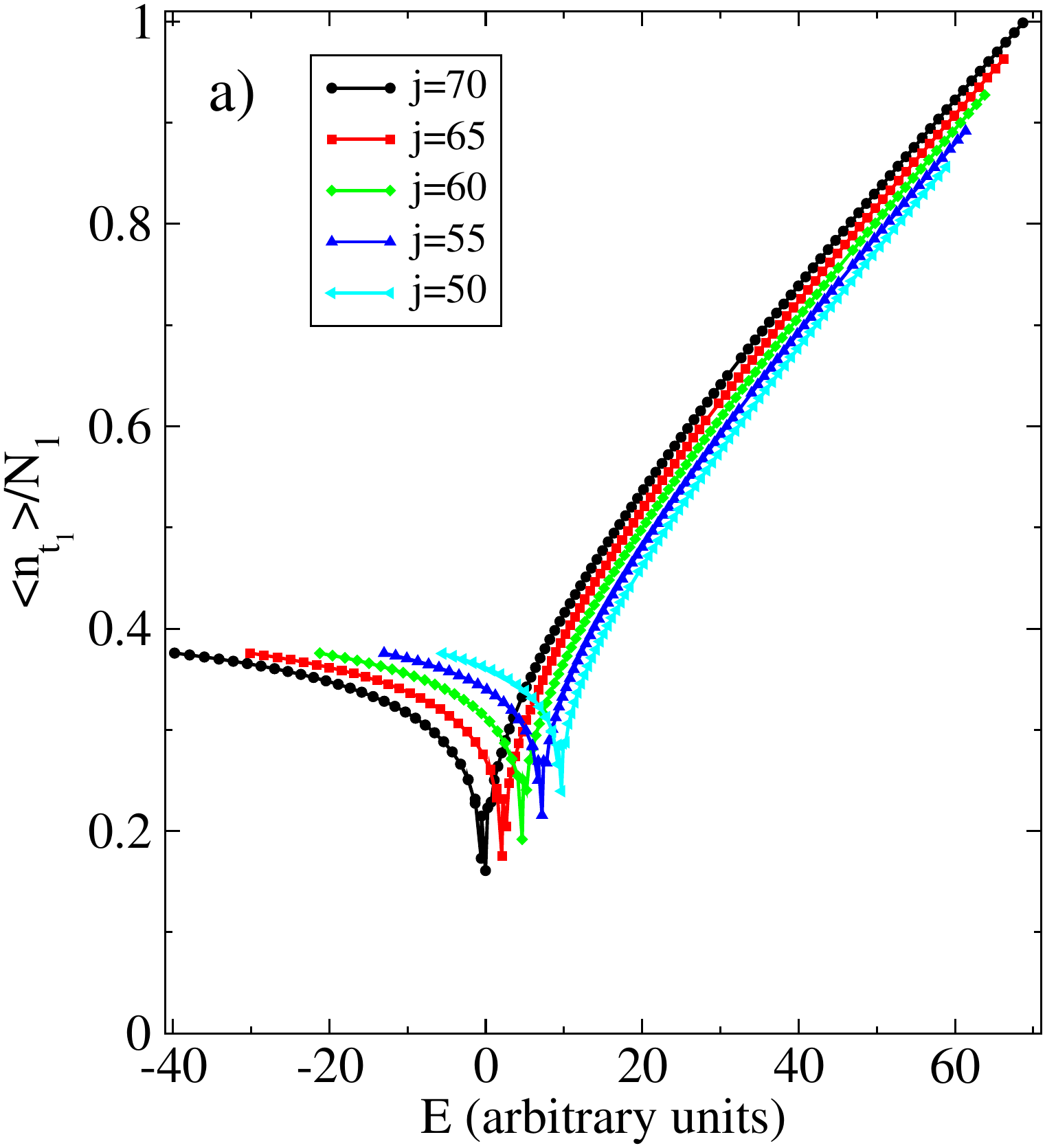}
&~~&
\includegraphics[width=0.40\linewidth]{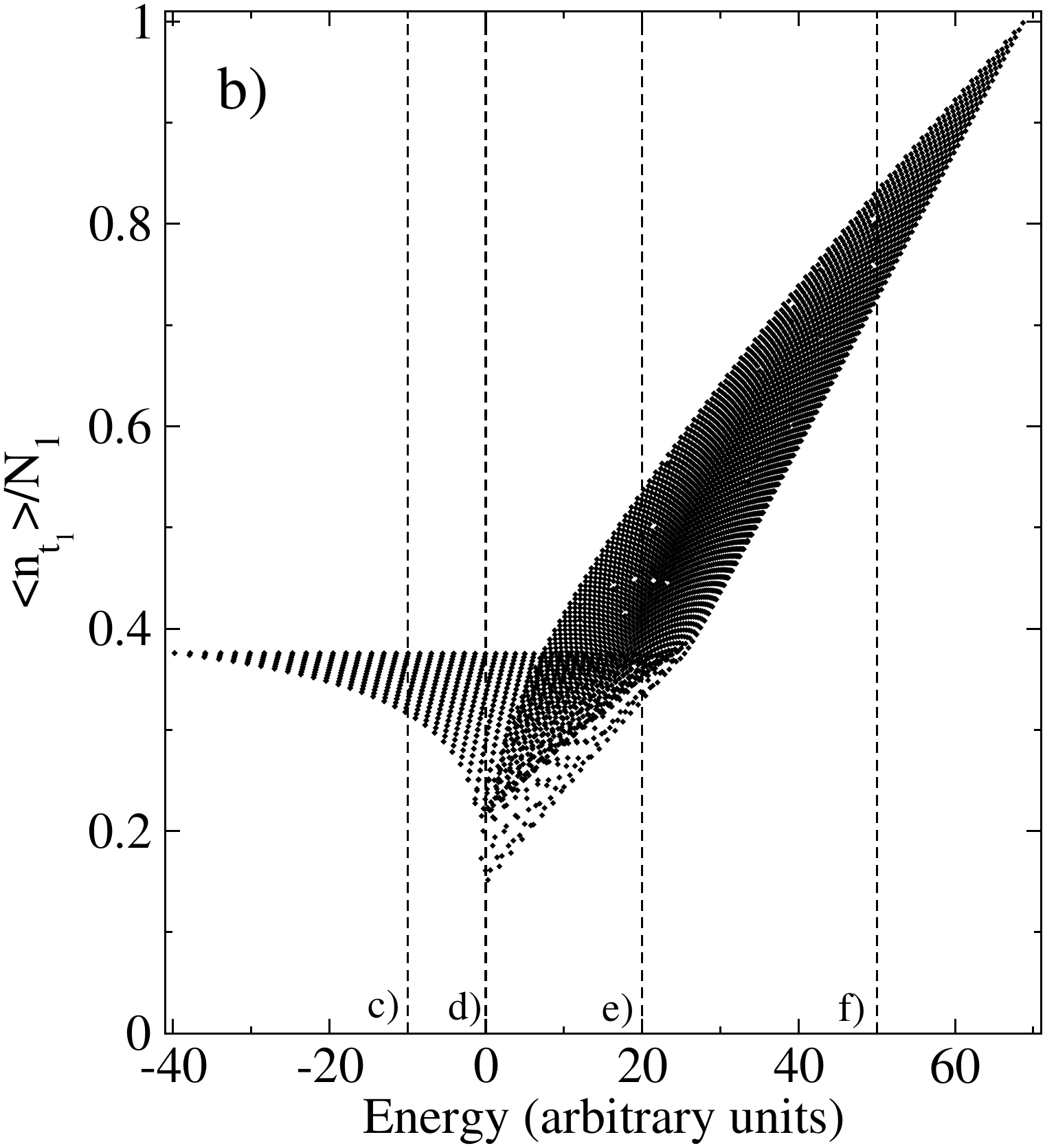}
\end{tabular}
\includegraphics[width=0.90\linewidth]{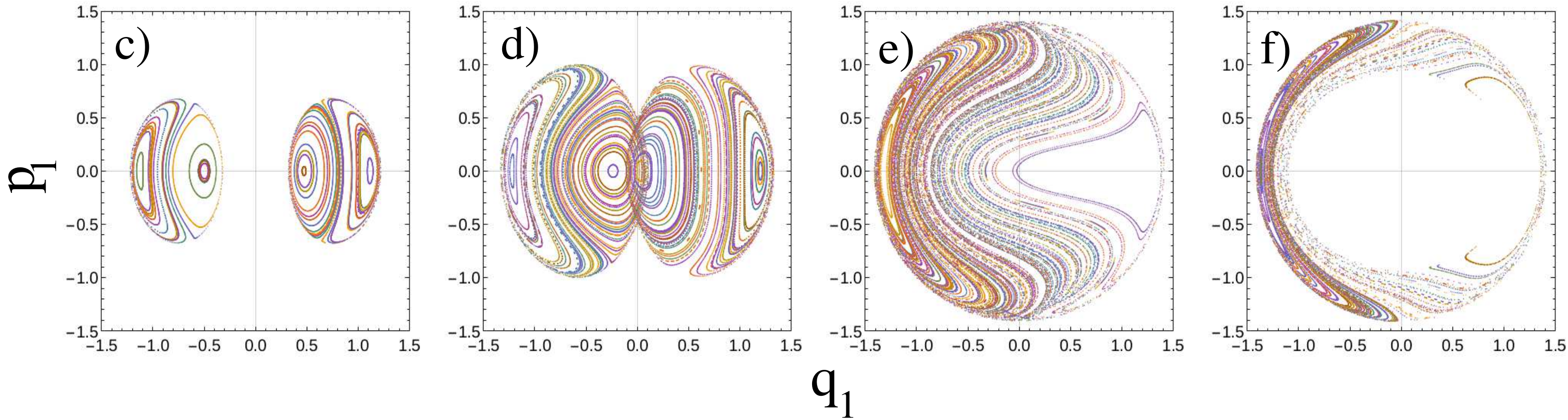}
\caption{Peres lattice for $\frac{\langle n_{t_1} \rangle}{N_1}$ versus the excitation
  energy for the same Hamiltonian and parameters as
  Fig.~\ref{fig-density-16}. a) For selected values of $j$, b) for all
  the states without distinction in $j$. Poincar\'e sections for
  energies $E=-10$, $E=0$, $E=20$, and $E=50$, in panels c), d), e),
  and f), respectively (energies marked in panel (b)).}   
\label{fig-n1-16}
\end{figure}

In Fig.~\ref{fig-n1-16} we plot the Peres lattice for $\frac{\langle
n_{t_1} \rangle}{N_1}$ as a function of the excitation energy: in
panel (a) the plot is done for each $j$ (different color for each $j$)
while in panel (b) all the states (all $j's$) are considered
together. In the latter case one can observe an almost global
regularity, but taking into account what is observed in panel (a), one
can notice that such a regularity is associated with the presence of
$j$ as an extra quantum number. The main feature of each set of
points, corresponding to given $j-$values, is the sharp dip observed
at energies around $E\approx 0$, with a shift in the dip position
towards higher energies as $j$ decreases, in a similar way to
Fig.~\ref{fig-density-16}a. Therefore, Fig.~\ref{fig-n1-16}b is simply
the superposition of different curves with sharp dips that move from
$E\approx 0$ to $E\approx 30$. Both below and above the ESQPT, the
Peres lattice shows an ordered pattern, pointing to a non-chaotic
behaviour. Note that in this case, we do not analyze the NNSD because
we have to separate in sets with the same value of $j$ and, therefore,
the number of states will be too low to calculate a reliable value of
$\eta$ (\ref{eta}).
% XXXXXXXXXXXXXXXXXXXXXXXX
However, one can calculate the Poincar\'e
sections to study the regularity and chaos interplay. In panels (c),
(d), (e), and (f) of 
Fig.~\ref{fig-n1-16} we depict the Poincar\'e sections for energies $E=-20$, $E=0$,
$E=30$, and $E=50$, respectively. Because of the underlying $U_{12}(2)$ symmetry, all the
cases correspond to a regular regime.

\begin{figure}
\begin{tabular}{ccc}
\includegraphics[width=0.40\linewidth]{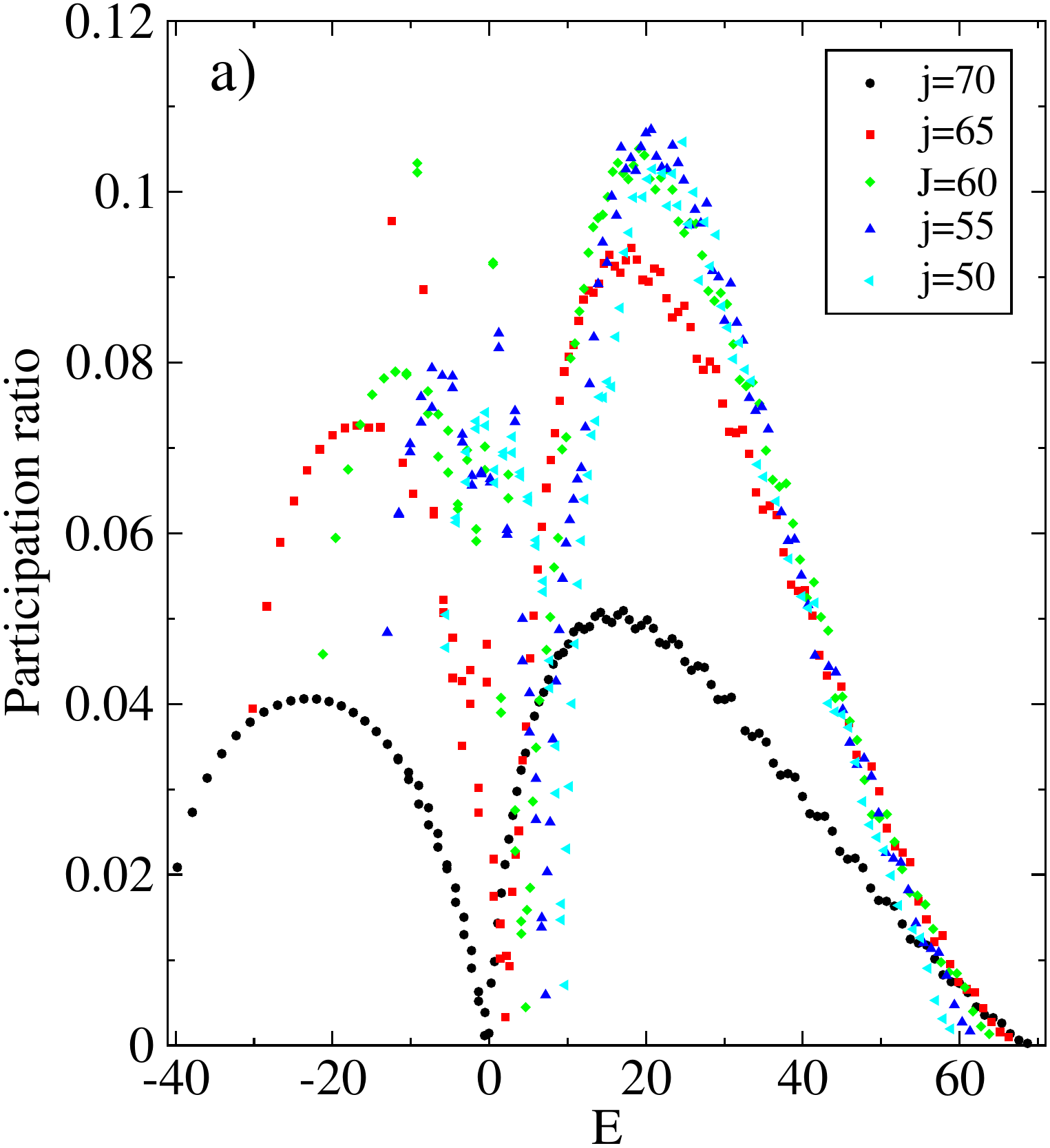}
&~~&
\includegraphics[width=0.40\linewidth]{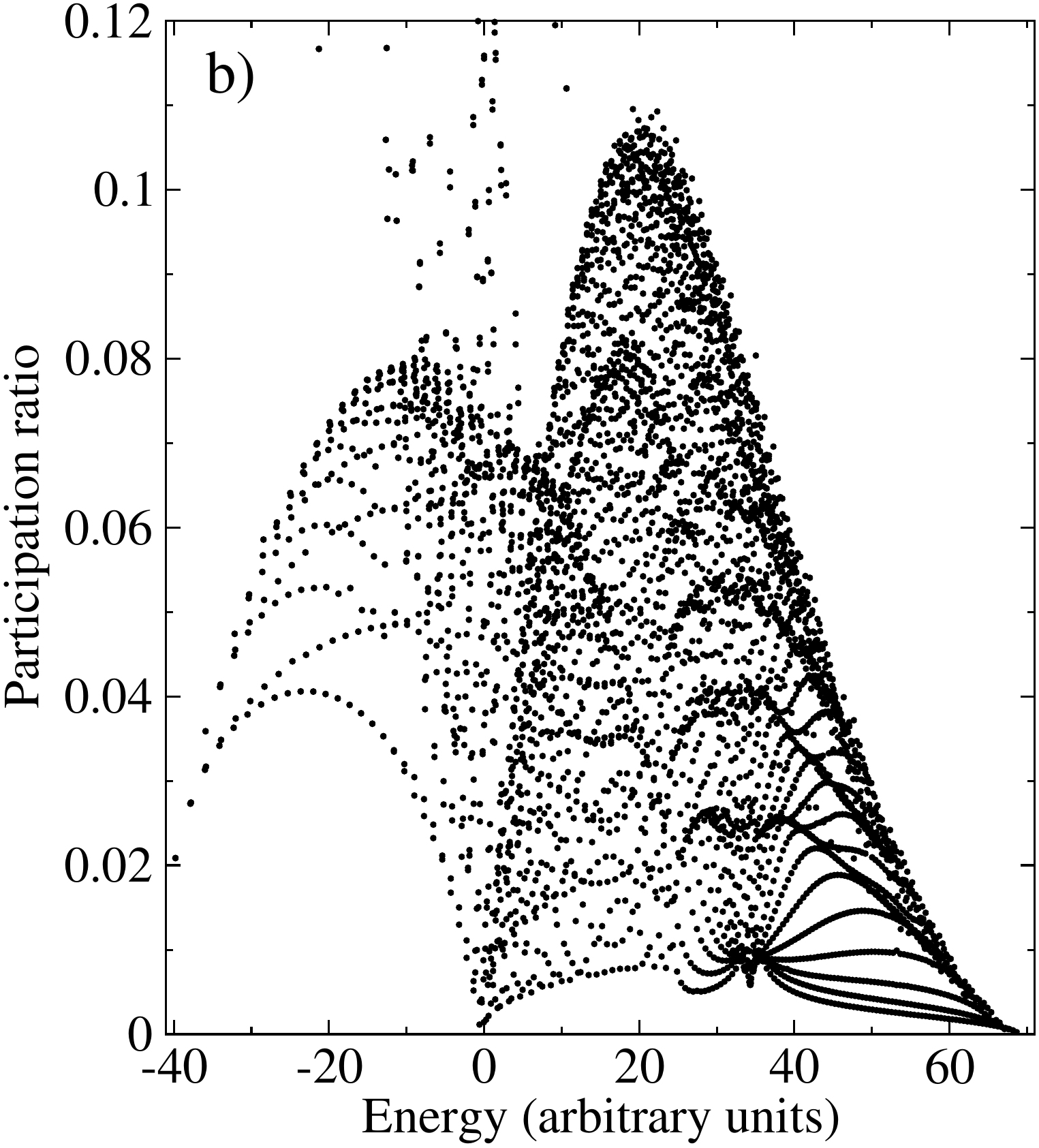}
\end{tabular}
\caption{Normalized participation ratio as a function of the excitation energy
  for the same Hamiltonian and parameters as in
  Fig.~\ref{fig-density-16}. a) For selected values of $j$, b) for all
  the states without distinction in $j$.}   
\label{fig-pr-16}
\end{figure}

In Fig.~\ref{fig-pr-16} we plot the value of the participation ratio as
a function of the excitation energy. Once more, in panel (a) we separate in
families with given $j-$values (different colors) and in panel (b) we plot all the
states. In each family depicted in panel (a) one can see how the participation
ratio starts increasing and reaches a maximum, then decreases with a
sharp minimum, then shows a new maximum, and finally ends with a
minimum. This behaviour was already described 
in \cite{Sant15} and simply reflects the strong ``localization'' of the
wave function at the ESQPT. Once more, the position of the minimum
moves towards higher energies for decreasing  $j-$values as it is shown in
Figs.~\ref{fig-density-16} and \ref{fig-n1-16}. Panel (b) corresponds
to the superposition of the previously described curves which leads to
the presence of certain parabolic curves somehow blurred. The set of
well defined lines in the right bottom part of Fig.~\ref{fig-pr-16}b
corresponds to curves with small values of $j$.  

%17
%\subsubsection{$y=1$, $y'=0$ ($y_1=1$, $y_2=1$), $x=1/2$}
%\label{sec-17}

The second selected case of interest corresponds to parameters $y=1$,
$y'=0$ ($y_1=1$, $y_2=1$), this is a line in the base of the phase
diagram going from $u_{12}(1)$  to the deformed region crossing the
line of first order phase transition which is located at $x_c\approx
4/5$ (point B in Fig.~\ref{fig-dia}). In particular we have selected
the point $x=1/2$. For this selection of the control parameters, the
system ground state 
is deformed and since $N_1=N_2$, the indexes $1$ and
$2$  can be interchanged, $u_{12}(2)$ is the dynamical algebra of the
Hamiltonian and $j$ is a good quantum number. Therefore, we are in a
similar situation to the case previously discussed but now, besides
the ESQPT, a set of states associated with the second (local) minimum
of the potential energy surface will appear. 

\begin{figure}
\begin{tabular}{ccc}
\includegraphics[width=0.40\linewidth]{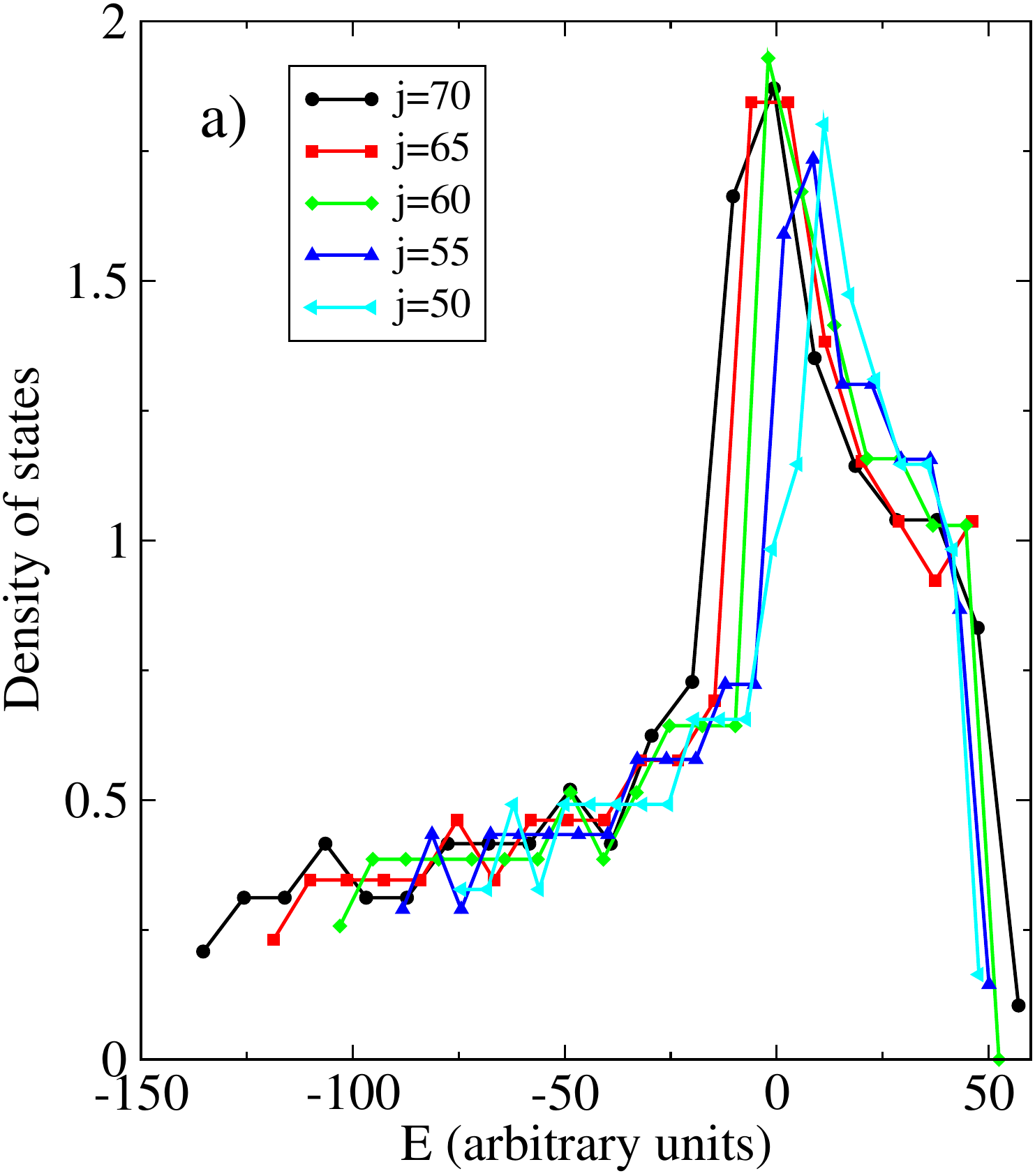}
&~~&
\includegraphics[width=0.40\linewidth]{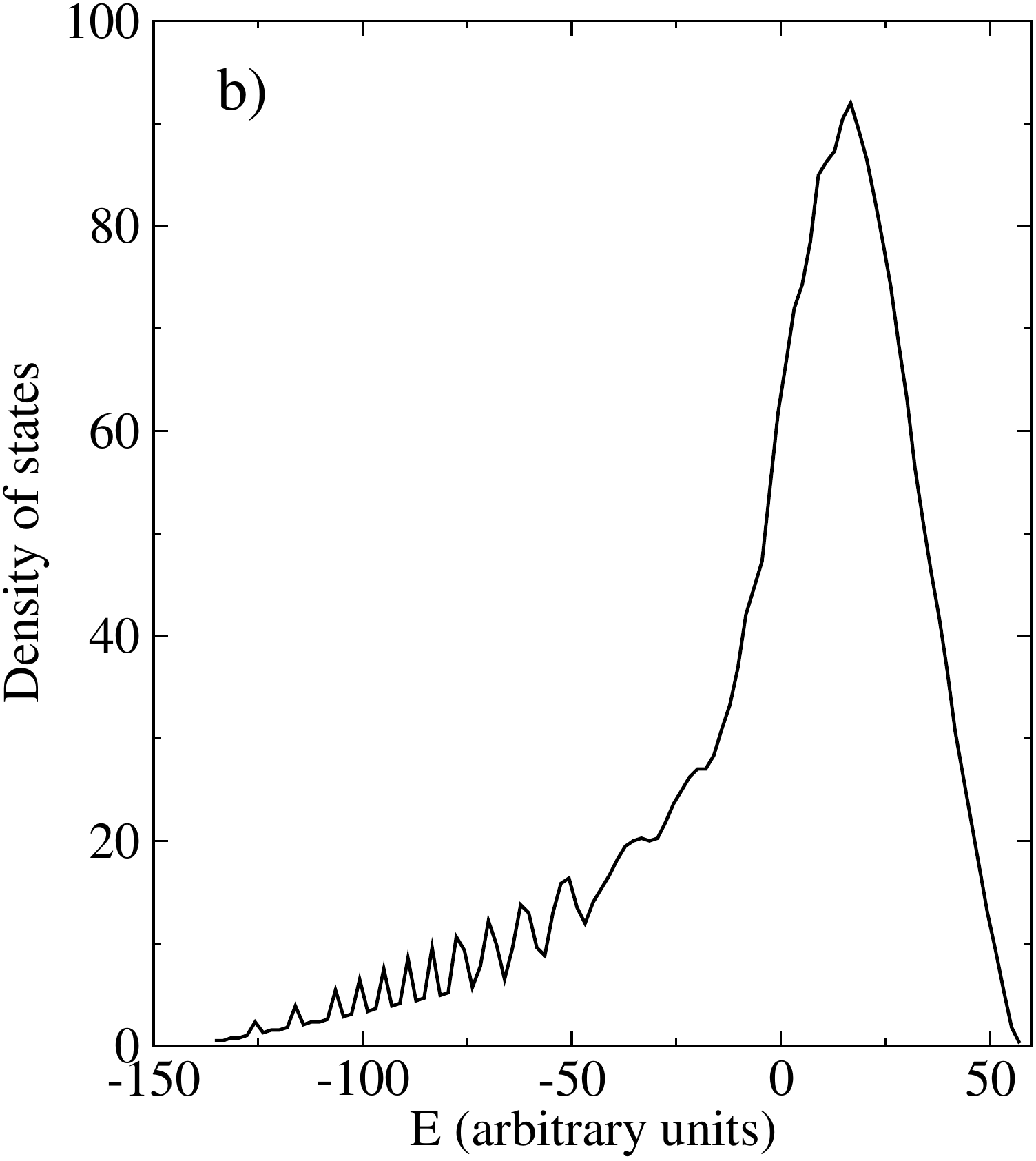}
\end{tabular}
\caption{Density of states as a function of the excitation energy 
  for a Hamiltonian with parameters
  $y=1$, $y'=0$, and $x=1/2$ and a number of bosons $N_1=N_2=70$. a)
  For selected values of $j$, b) for all the states without
  distinction in $j$.}  
\label{fig-density-17}
\end{figure}

We start with the analysis of the density of states. In
Fig.~\ref{fig-density-17}, the density of states are plotted versus
the excitation energy. In panel (a) each color line corresponds to a
given $j-$value, while in panel (b) the total density of states is
plotted independently of the $j-$values. In panel (a), in addition to
the peak close to zero that is the precursor of the lambda divergence,
a previous finite discontinuity is observed very close in energy
(negative). In panel (b), a relatively broad peak is observed above
zero energy, but nothing qualitatively different to
Fig.~\ref{fig-density-16}b which corresponds to an ESQPT without a
local minimum in the potential energy surface.

\begin{figure}
\begin{tabular}{ccc}
\includegraphics[width=0.40\linewidth]{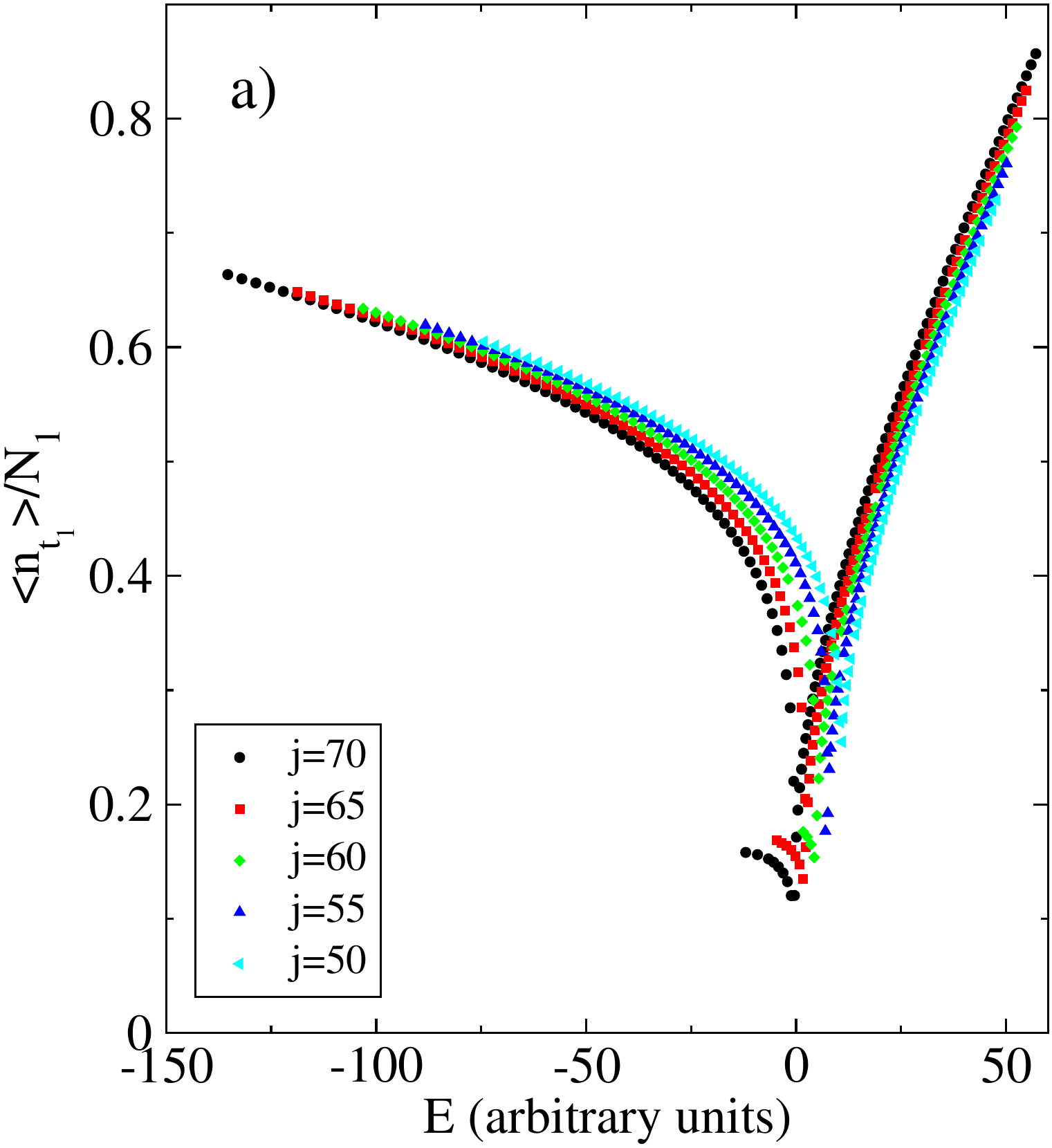}
&~~&
\includegraphics[width=0.40\linewidth]{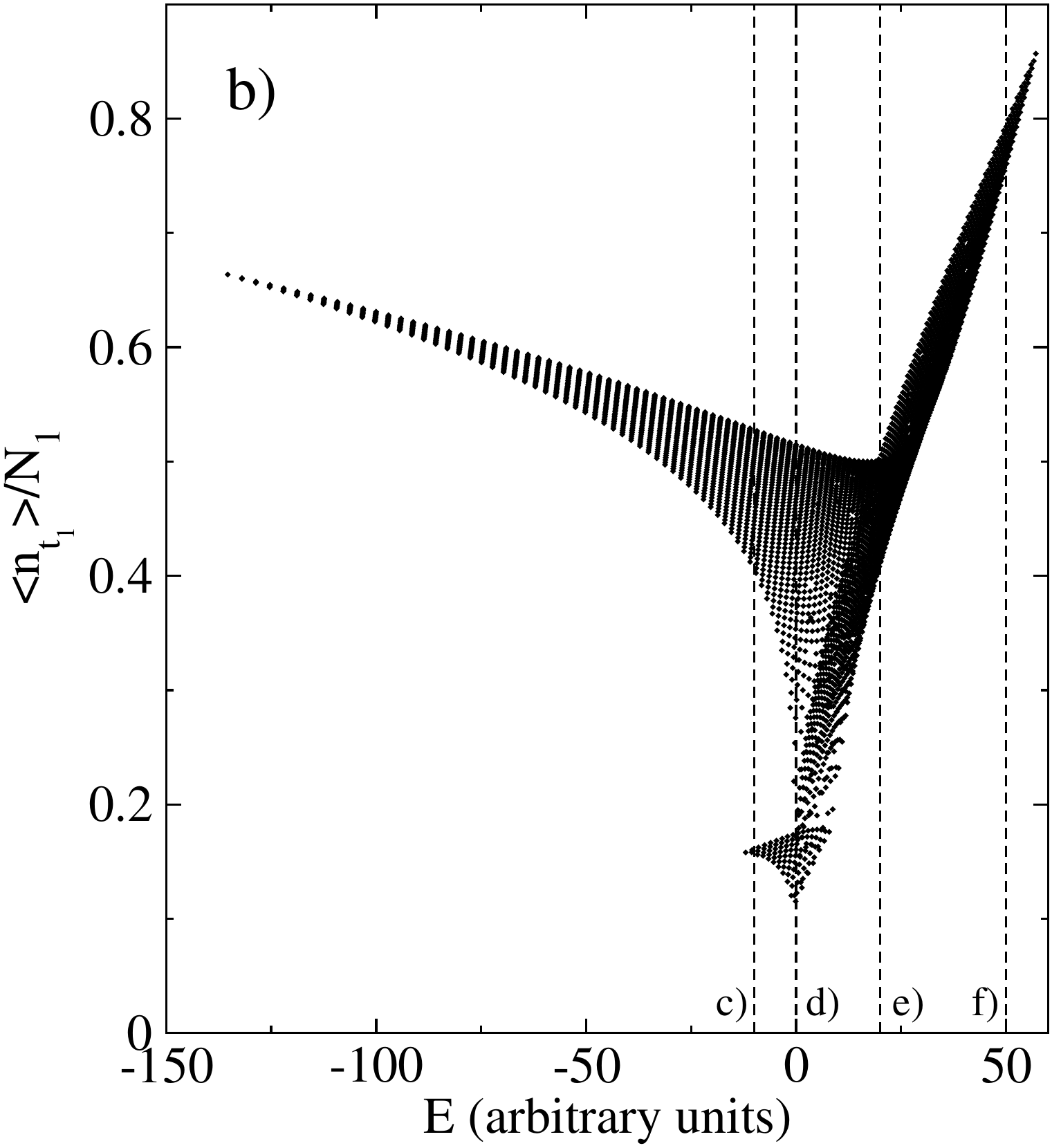}
\end{tabular}
\includegraphics[width=0.90\linewidth]{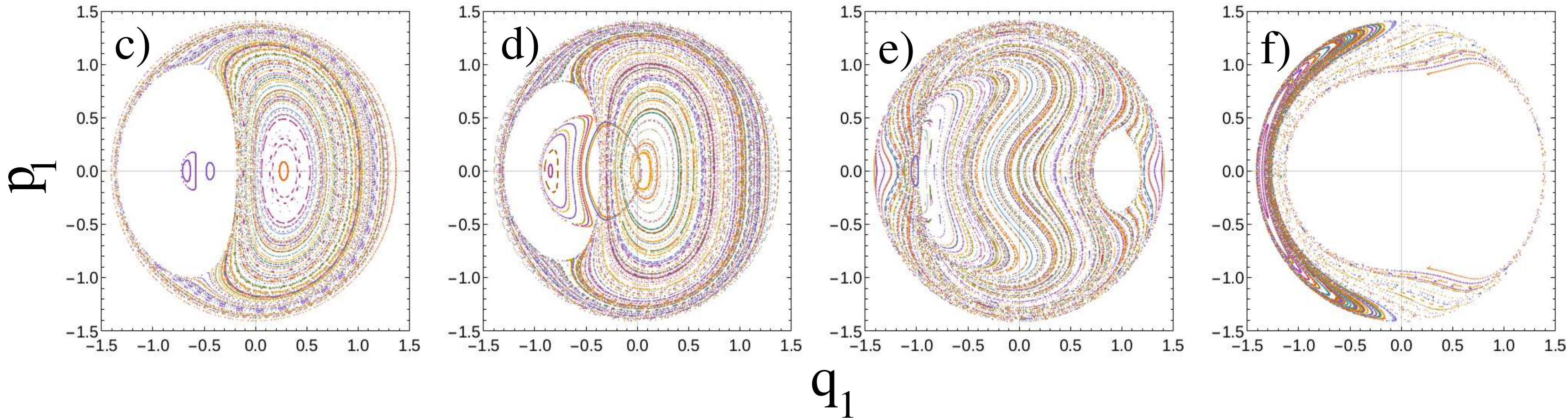}
\caption{Peres lattice for $\frac{\langle  n_{t_1} \rangle}{N_1}$ versus the excitation
  energy for the same Hamiltonian and parameters than in
  Fig.~\ref{fig-density-17}. a) For selected values of $j$, b) for all
  the states without distinction in $j$. Poincar\'e sections for
  energies $E=-10$, $E=0$, $E=20$, and $E=50$, in panels c), d), e),
  and f), respectively (energies marked in panel (b)).}   
\label{fig-n1-17}
\end{figure}

In Fig.~\ref{fig-n1-17} we present the Peres lattice for
$\frac{\langle n_{t_1} \rangle}{N_1}$ versus the excitation energy for
given $j-$values (different colors) in panel (a), or for the whole set
of 
states in panel (b). Here, narrow dips are observed in panel (a) 
around the energy of the ESQPT. The main difference with respect to
the previous case is the appearance of a second family of states (see
lowest part of Fig.~\ref{fig-n1-17} in both panels)
related with the local minimum of the potential energy surface.  
Note that also in this case, we do not analyze the NNSD because
we have to separate in sets with the same $j-$value and,
therefore, the states will be too few to calculate the value of
$\eta$.
%XXXXXXXXXXXXXXXXXXXXX
As in previous case, one can calculate the Poincar\'e
sections to study the regularity and chaos interplay.  In panels (c),
(d), (e), and (f) of  Fig.~\ref{fig-n1-17} we depict the Poincar\'e
sections for energies $E=-10$, $E=0$, $E=20$, and $E=50$,
respectively. Because of the underlying $U_{12}(2)$ symmetry, all the 
cases correspond to a regular regime, too.

\begin{figure}
\begin{tabular}{ccc}
\includegraphics[width=0.40\linewidth]{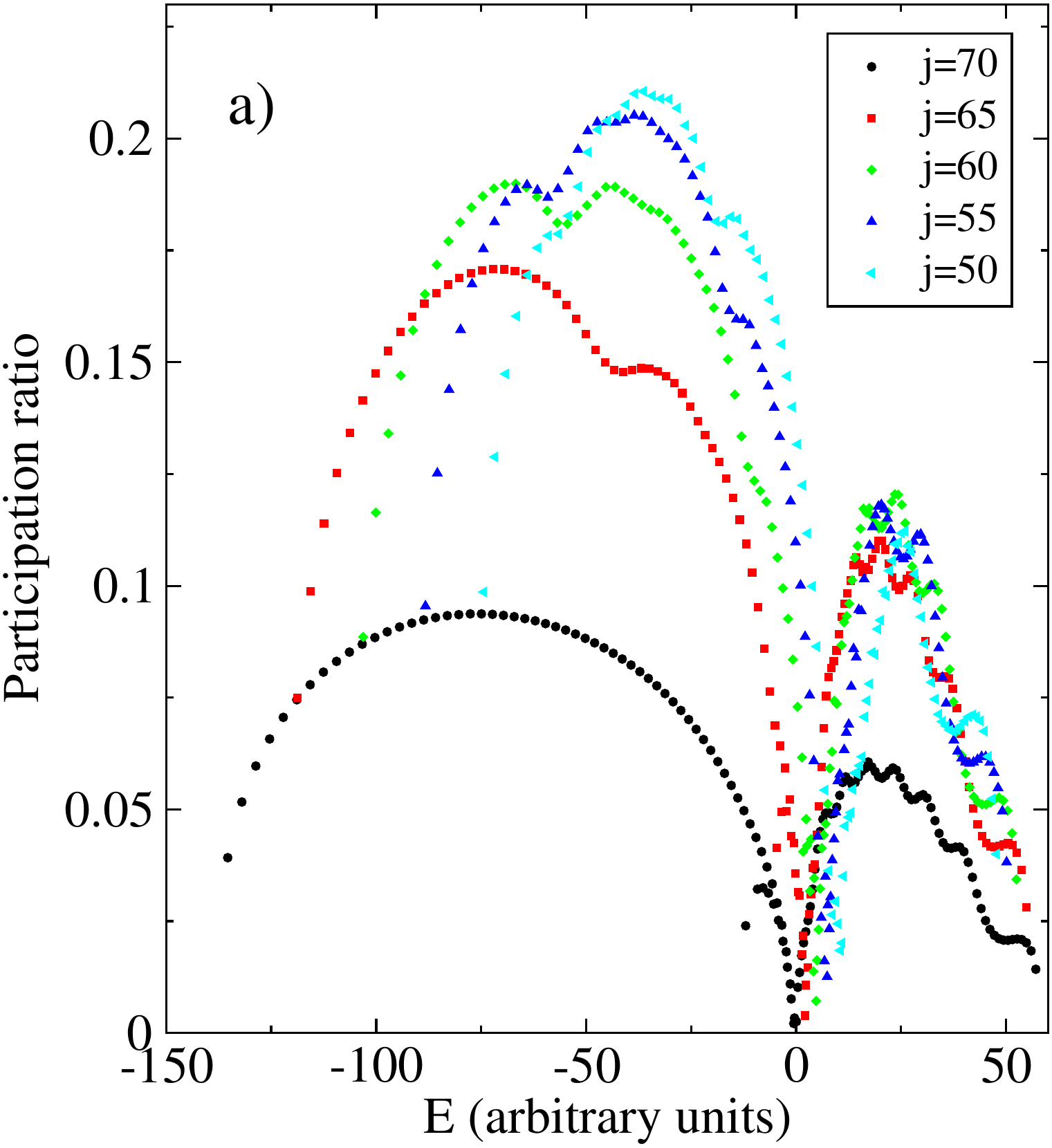}
&~~&
\includegraphics[width=0.40\linewidth]{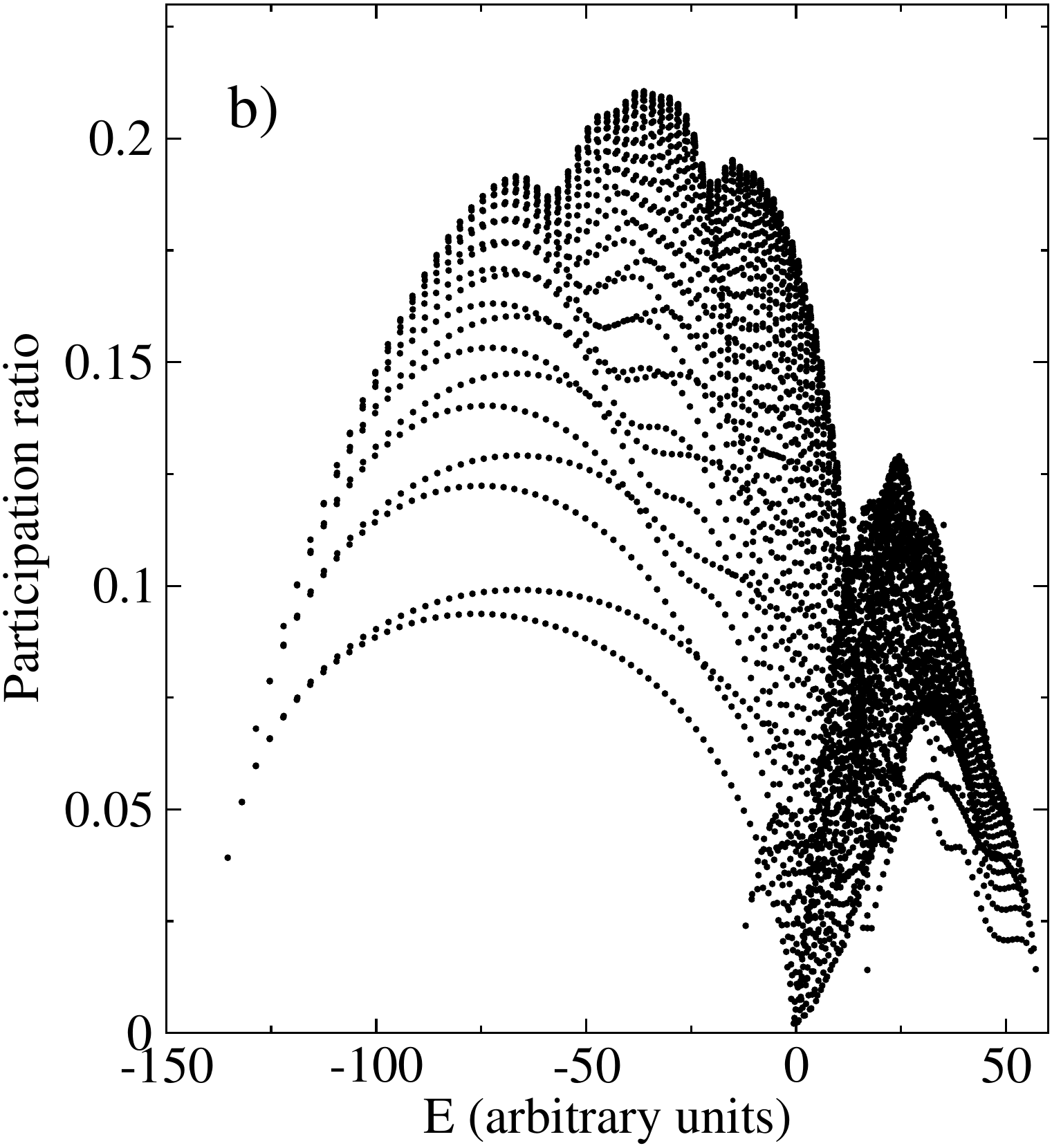}
\end{tabular}
\caption{Participation ratio as a function of the excitation energy 
  for the same Hamiltonian and parameters as in
  Fig.~\ref{fig-density-17}. a) For selected $j-$values, b) for all
  the states without distinction in $j$.}   
\label{fig-pr-17}
\end{figure}

In Fig.~\ref{fig-pr-17} we depict the participation ratio value as a
function of the excitation energy,  
for selected particular $j-$values (different colors) in panel (a),
and for the whole set of states in panel (b). In panel (a) each family
of states has a narrow minimum around zero energy with broad maxima at
right and left. Here, for each $j-$family a second set of states,
related with the local maximum of the energy surface, appears just below
the energy of the ESQPT, but it can be almost unnoticed. It is worth
to mention that even in panel (b) the position of the ESQPT critical
energy is clearly marked with a relatively narrow dip. 

Till now we have only studied cases where $u_{12}(2)$ was the
dynamical algebra of the Hamiltonian and, therefore, they essentially
correspond to the superposition of several one-fluid Lipkin systems
with $2j$ number of bosons.  Now we will move to the more interesting
$y_1\neq y_2$ cases, where $j$ is no longer a good quantum number. The
already studied cases have taught us how the patterns observed in
Figs. \ref{fig-density-16}b, \ref{fig-n1-16}b, \ref{fig-pr-16}b,
\ref{fig-density-17}b, \ref{fig-n1-17}b, and \ref{fig-pr-17}b are a
consequence of the superposition of lines corresponding to different
$j-$values, in other words, a consequence of the underlying
$u_{12}(2)$ symmetry.

\begin{figure}
\begin{tabular}{ccc}
\includegraphics[width=0.30\linewidth]{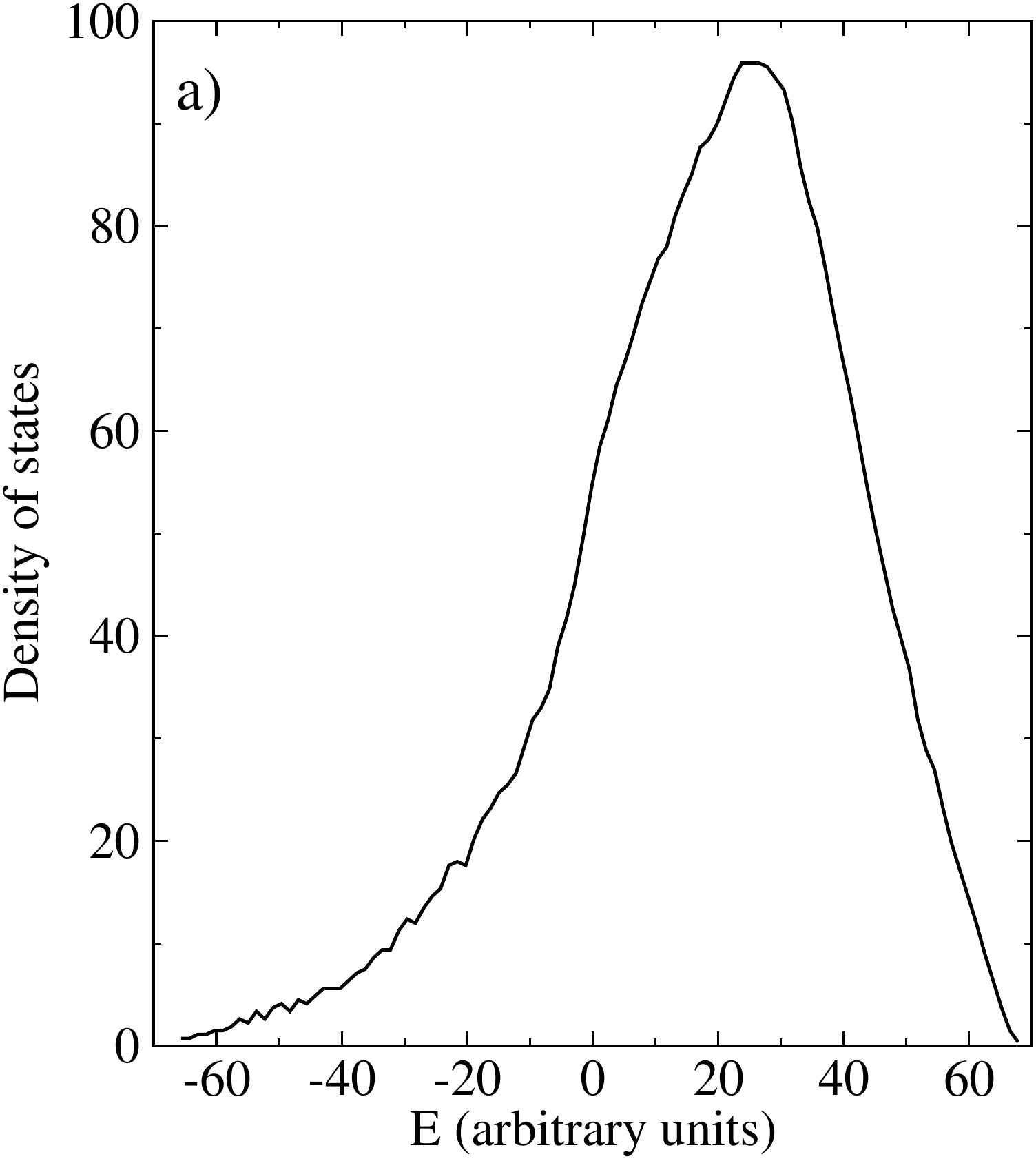}
&
\includegraphics[width=0.30\linewidth]{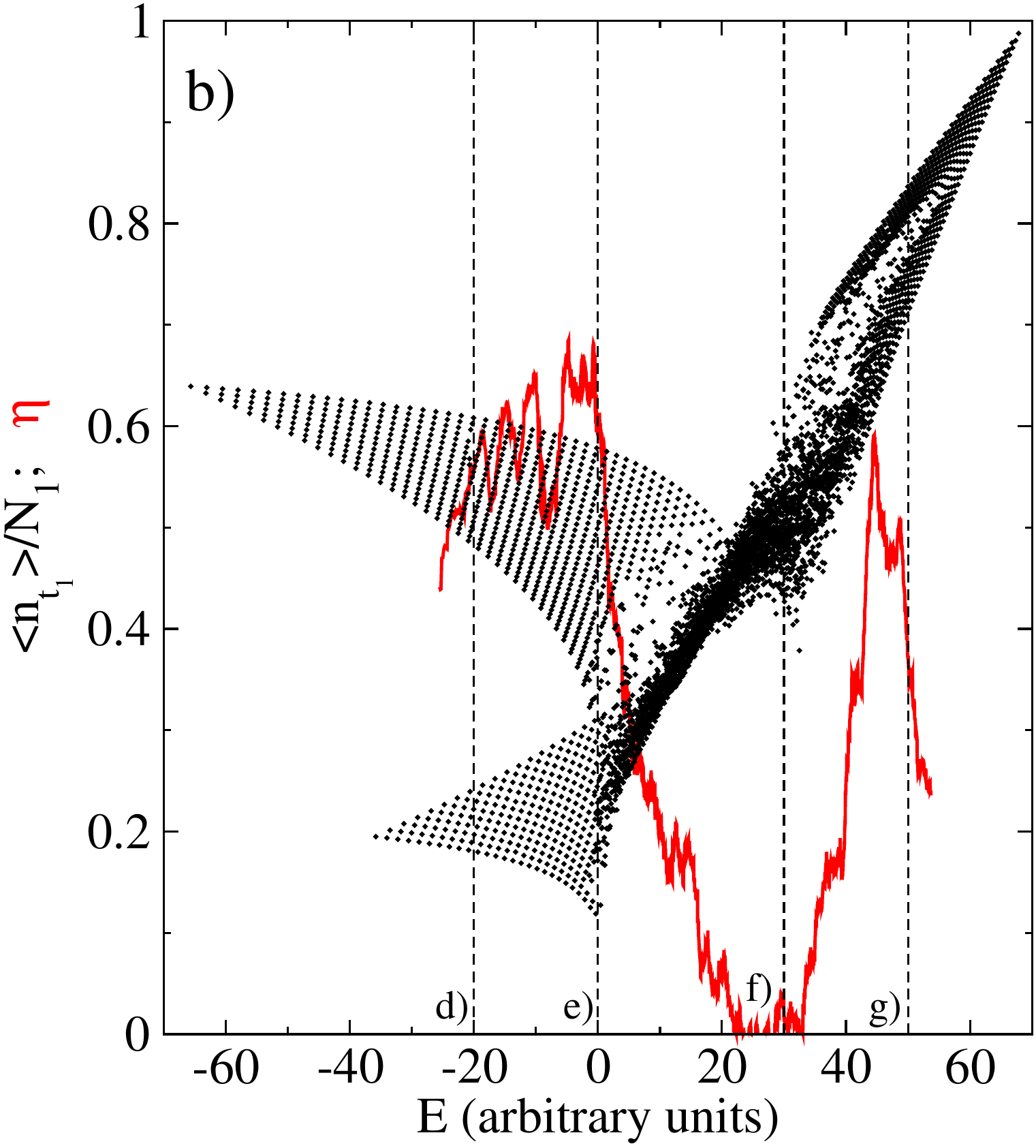}
&
\includegraphics[width=0.30\linewidth]{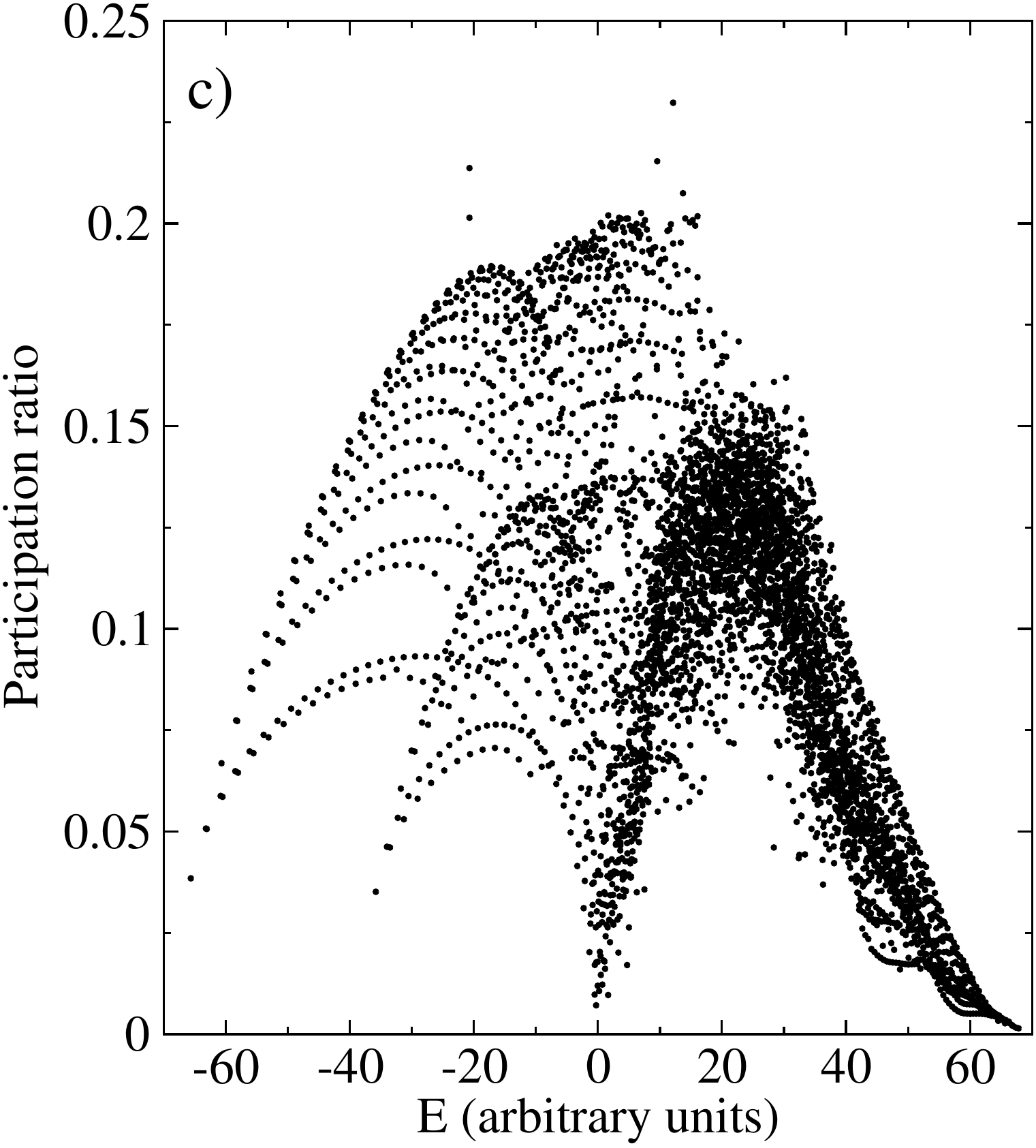}
\end{tabular}
\includegraphics[width=.95\linewidth]{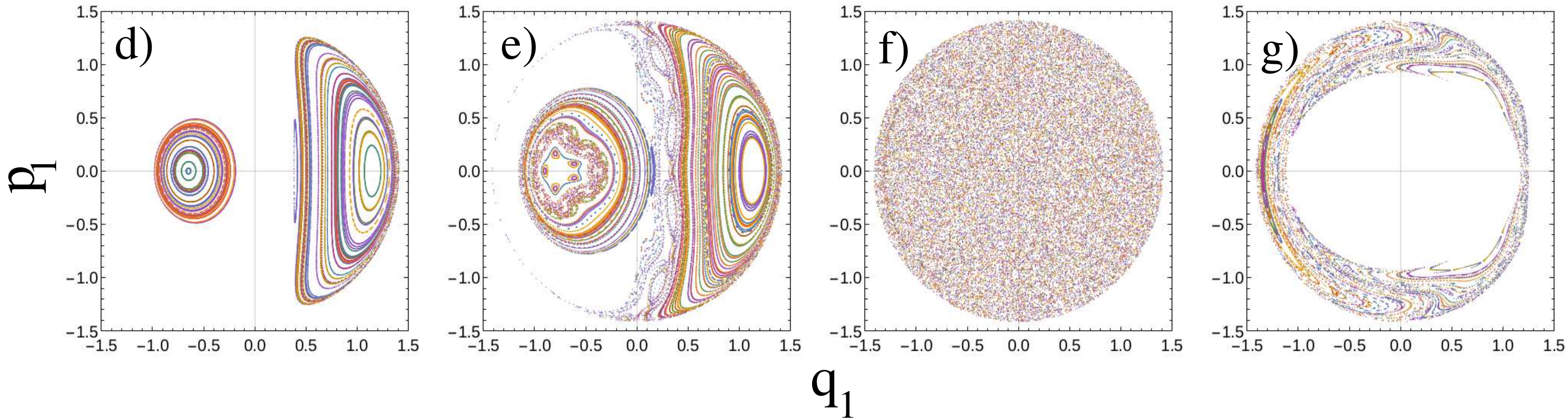}
\caption{Different observables as a function of the excitation energy 
  for a Hamiltonian with parameters $y=1/4$, $y'=3/4$, and $x=1/2$ and
  a number of bosons $N_1=N_2=70$: a) Density of states, b) Peres
  lattice (black points) for $\frac{\langle  n_{t_1} \rangle}{N_1}$
  and $\eta-$value (red solid curve), c) participation ratio, 
%XXXXXXXXXXXXXXXXXXXXXXXXX
  d), e), f), and g) 
  Poincar\'e sections for energies $E=-20$, $E=0$, $E=30$, and $E=50$,
  respectively (energies marked in panel (b)).}    
\label{fig-comb-21}
\end{figure}

Next case of interest corresponds to the parameter values $y=1/4$,
$y'=3/4$ ($y_1=1$, $y_2=-1/2$) which is a point in the line from
spherical $u_{12}(1)$ to the deformed region but not in the base of
the phase diagram, and consequently does not preserve the $u_{12}(2)$
symmetry. This line crosses the first order surface located around
$x_c \approx 4/5$. We have selected the point $x=1/2$ which
corresponds to a deformed ground state (point C in
Fig.~\ref{fig-dia}). In this case, as already told, it is no longer
possible to separate the states in terms of $j$. The potential energy
surface of this Hamiltonian has a global deformed minimum and another
local deformed one separated by a maximum.  In
Fig.~\ref{fig-comb-21}a, the density of states is plotted as a
function of the excitation energy. It can be observed that this
quantity is not marking correctly the presence of the ESQPT at zero
energy, as already learnt from the preceding calculations. It is worth
noting that, although this case presents a second family of states
associated to the local minimum starting at $E\approx - 40$, nothing
is observed in the density of states. In Fig.~\ref{fig-comb-21}b the
Peres lattice for $\langle n_{t_1}\rangle/N_1$ is depicted as a
function of the excitation energy. This quantity shows three clear
regions, two on them at energies below the ESQPT critical energy (zero
energy) and the third one above. The two lowest families of points
correspond to states that are located in the well of the global
deformed minimum ($\langle n_{t_1}\rangle/N_1\approx 0.6$) and in the
well of the local one ($\langle n_{t_1}\rangle/N_1\approx 0.2$), also
deformed. These two regions show a very regular pattern. The third
region is above $E\approx 0$ and presents a relatively disordered
structure, except in particular regions, as is $0.8<\langle
n_{t_1}\rangle/N_1 <1$. The Peres lattice is clearly showing the
existence of two different deformed phases, non-symmetric, with a more
regular behaviour below zero energy and a symmetric phase with certain
degree of chaoticity above the energy of the ESQPT.
In panel (b) the value of $\eta$ (red line), representing the NNSD, as
a function of the energy is plotted too. This quantity shows a sudden
decrease at the energy of the ESQPT, therefore pointing to a spectrum
with a more regular behaviour below the ESQPT and more chaotic
above. However, even above the ESQPT, there is a regular region at
$E\approx 50$ with $\langle n_{t_1}\rangle/N_1\approx 0.7$. This
behaviour has been already observed in other models such as the
Bose-Hubbard Hamiltonian \cite{Rela14}.  Note that we do not reach the
whole range of energies because we exclude the 10\% of states with
lowest and highest energies to calculate $\eta$.
In Fig.~\ref{fig-comb-21}c it is depicted the participation ratio
versus the excitation energy. Here, one can note a rather different
structure for energies below and above the critical energy of the
ESQPT. For energies below zero, one can see two sets of inverted
parabolas, with a minimum around zero. The two sets correspond to the
two branches already seen in Fig.~\ref{fig-comb-21}b. For energies
above the ESQPT a single inverted parabola, though rather blurred, is
observed. This region presents a behaviour similar to the one obtained
with $u_{12}(2)$ conserving Hamiltonians. 
% XXXXXXXXXXXXX 
Finally, panels
(d), (e), (f) and (g) correspond to Poincar\'e sections with
energies $E=-20$, $E=0$, $E=30$, and $E=50$, respectively. These
figures confirm the latter statements, {\it i.e.}, that below the
ESQPT energy a regular behaviour exits, as panel (d) confirms, at the
ESQPT energy chaotic orbits star to appear (as shown in panel (e)),
and above the ESQPT energy, regions with chaotic (see panel (f)) and partially regular
motion coexist (see panel (g)).
%21
%\subsubsection{$y=1/4$, $y'=3/4$ ($y_1=1$, $y_2=-1/2$), $x=1/2$}
%\label{sec-12}

%\subsubsection{$y=1/4$, $y'=3/4$ ($y_1=1$, $y_2=-1/2$), $x=1/2$}
%\label{sec-12}

\begin{figure}
\begin{tabular}{ccc}
\includegraphics[width=0.30\linewidth]{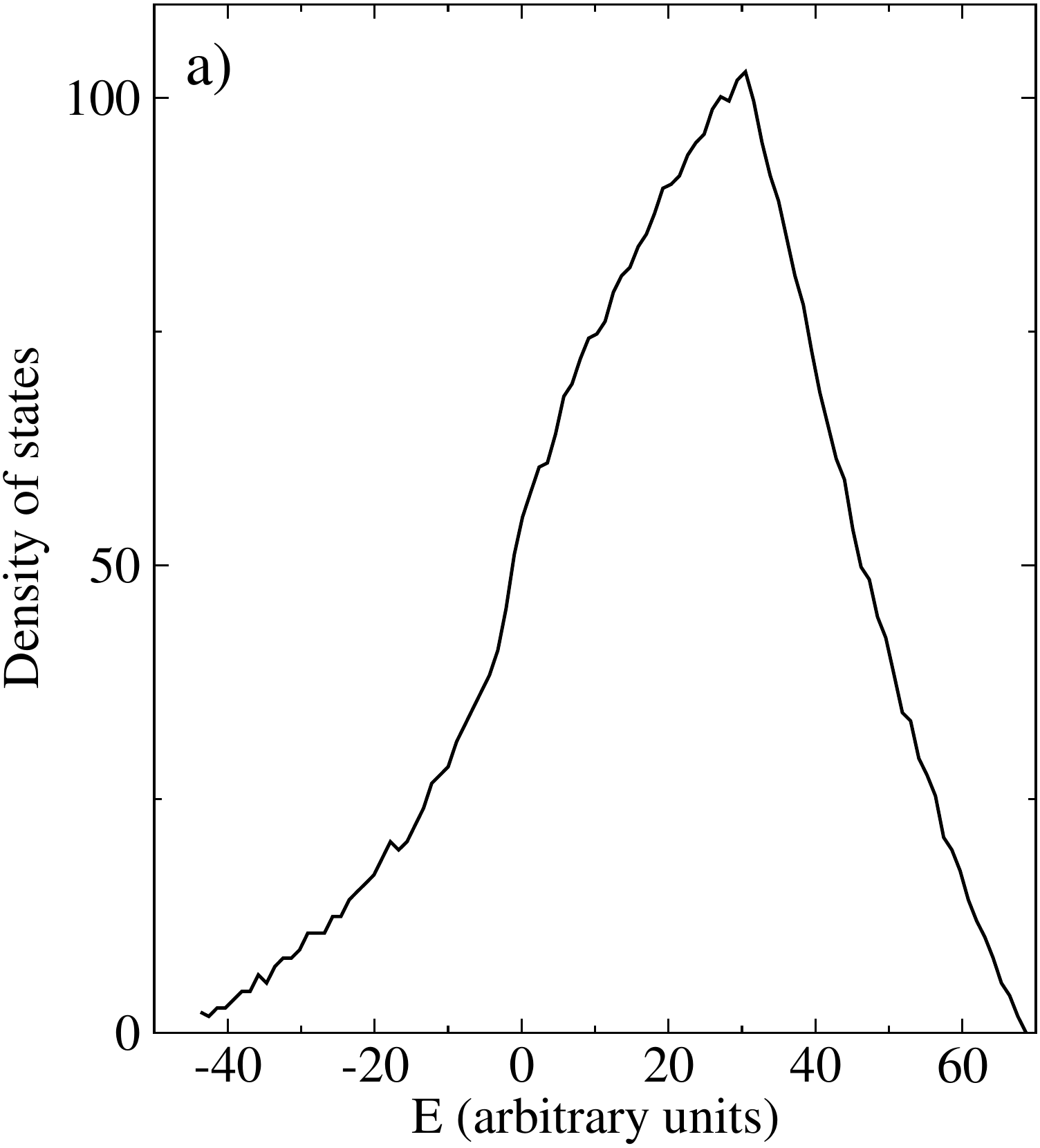}
&
\includegraphics[width=0.30\linewidth]{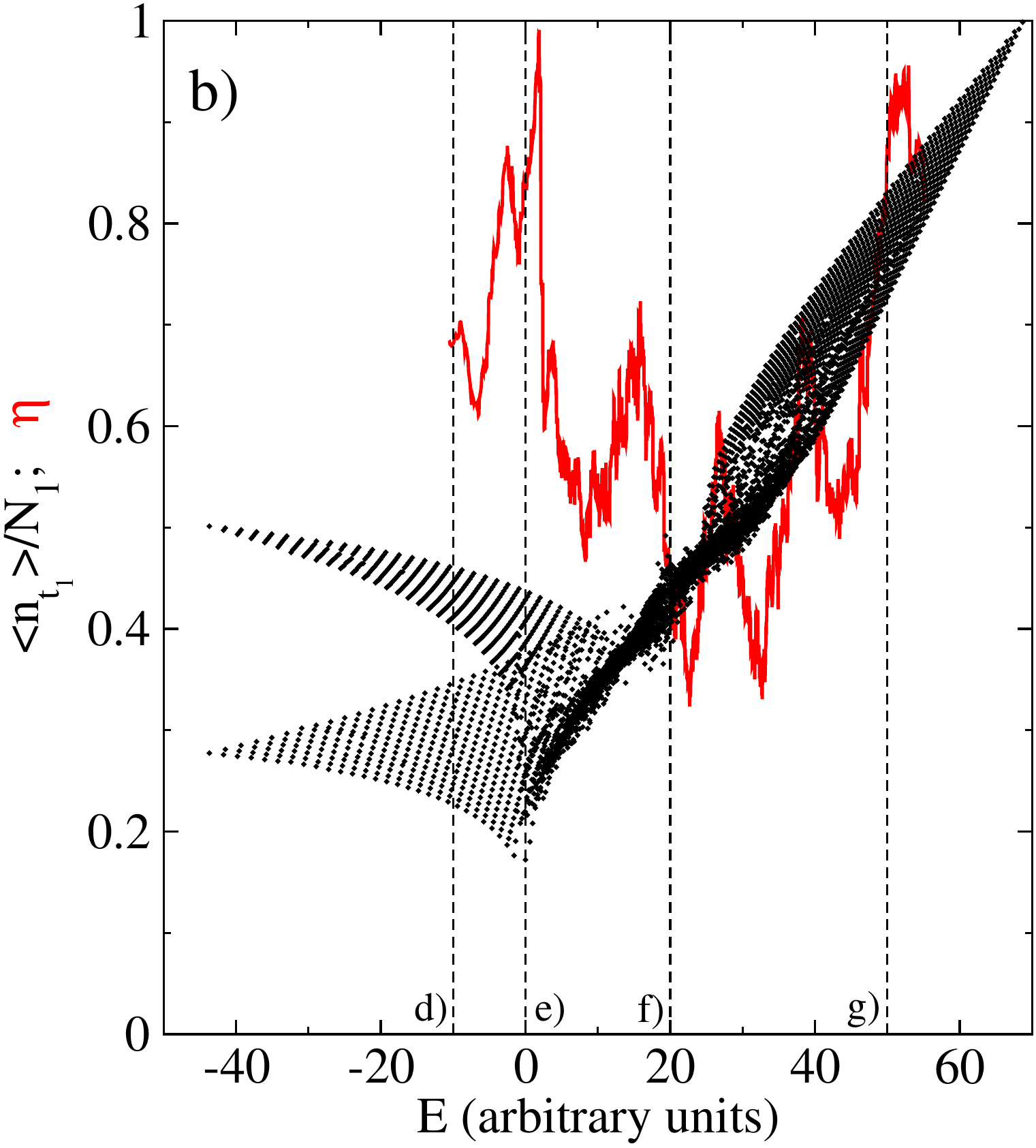}
&
\includegraphics[width=0.30\linewidth]{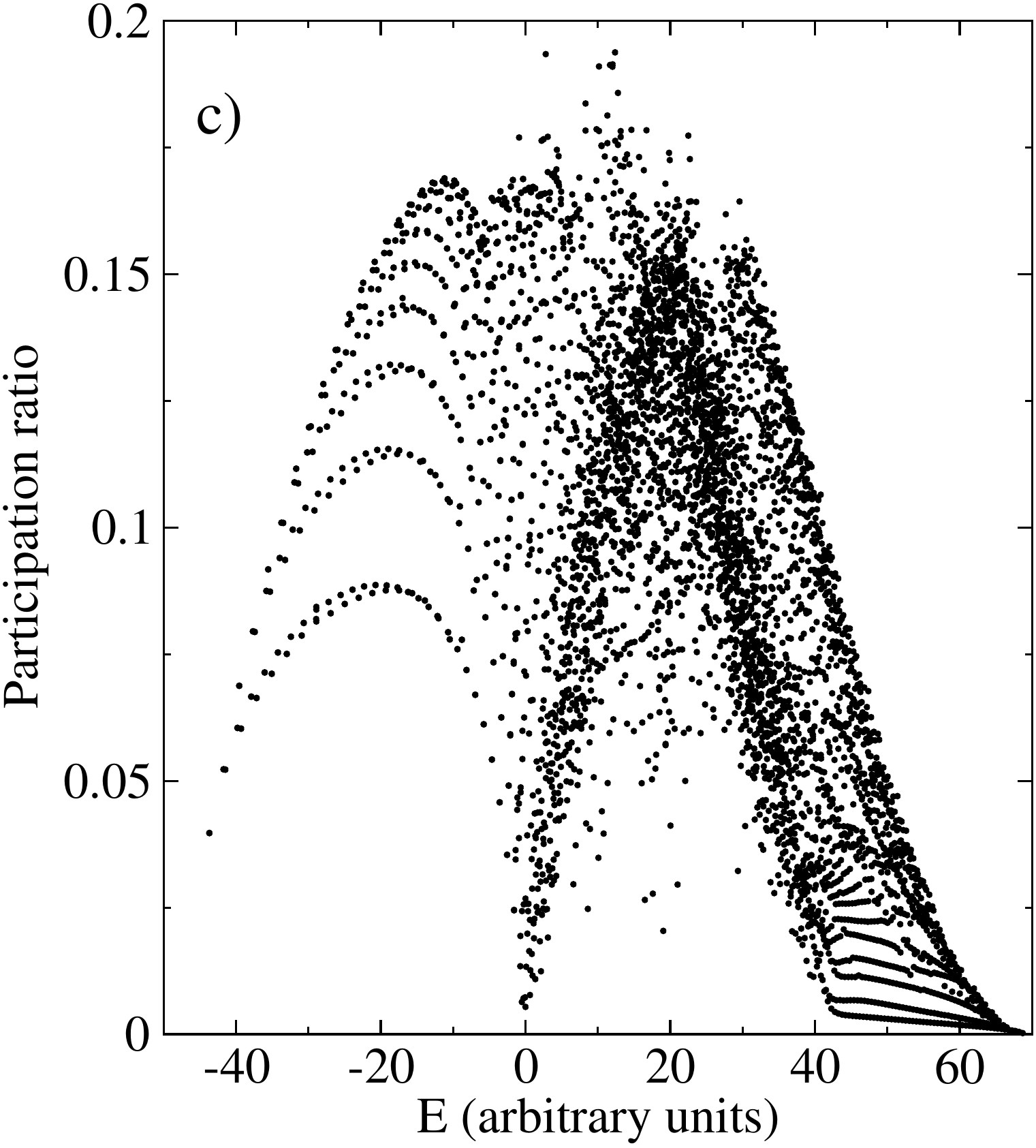}
\end{tabular}
\includegraphics[width=1\linewidth]{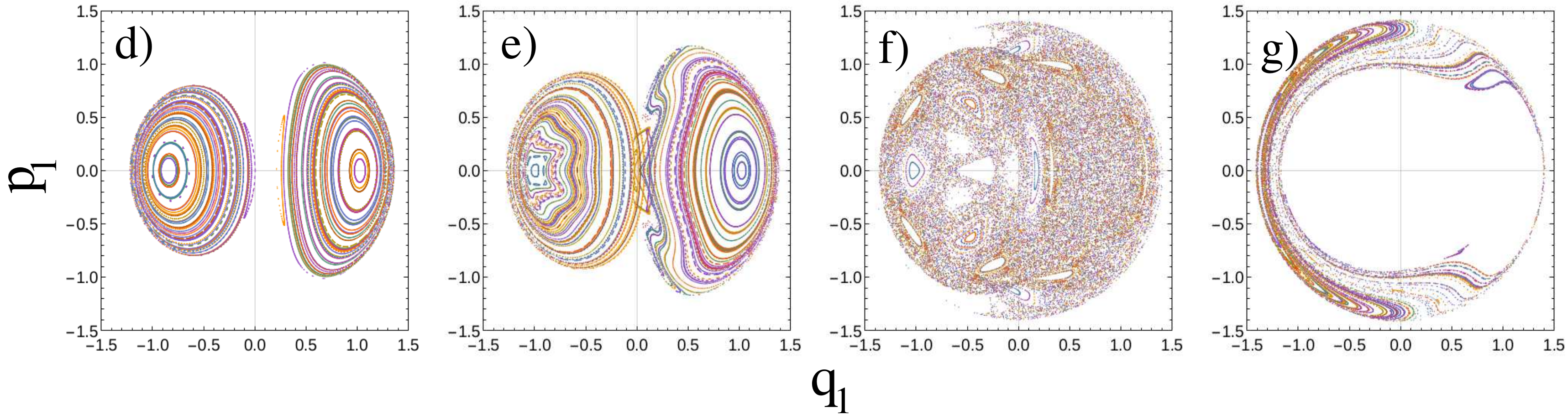}
\caption{Same as in Fig.~\ref{fig-comb-21} but for $y=0$, $y'=1/2$,
  and $x=1/2$. 
%XXXXXXXXXXXXXXXXXXXXXXXXXXXXXXXXXXX
  Panels d), e), f), and g) are
  Poincar\'e sections for energies $E=-10$, $E=0$, $E=20$, and $E=50$.}  
\label{fig-comb-13}
\end{figure}

\begin{figure}
\begin{tabular}{ccc}
\includegraphics[width=0.30\linewidth]{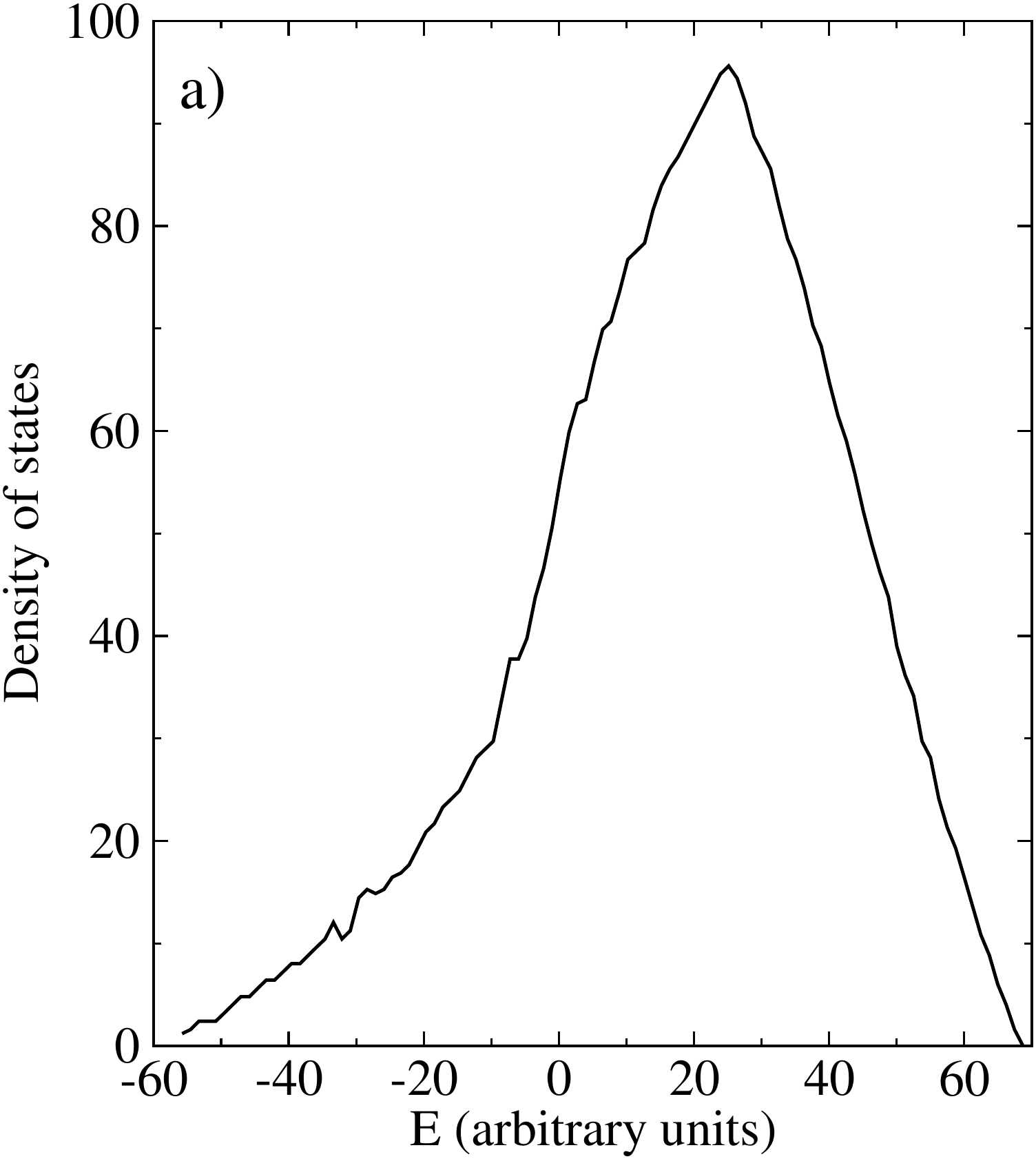}
&
\includegraphics[width=0.30\linewidth]{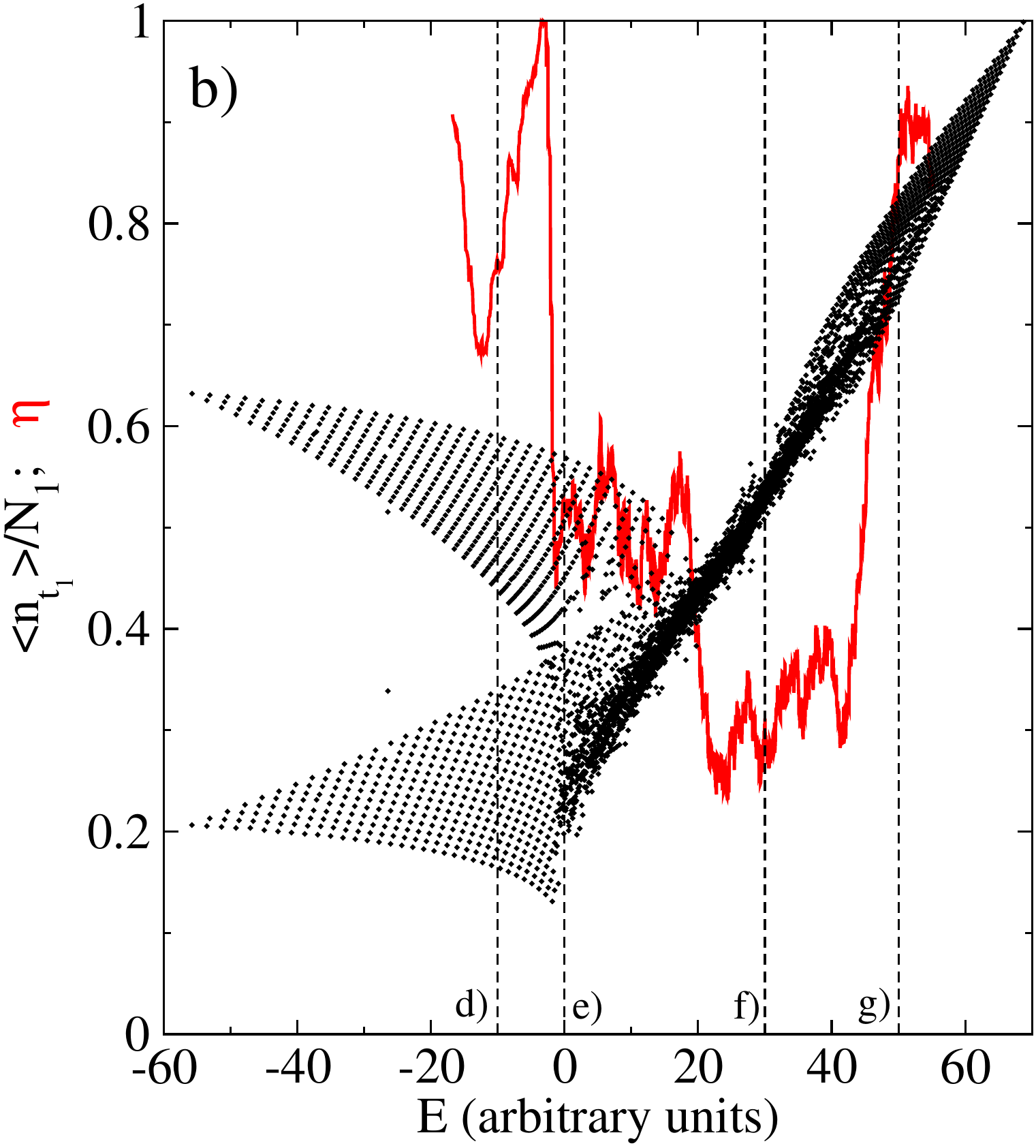}
&
\includegraphics[width=0.30\linewidth]{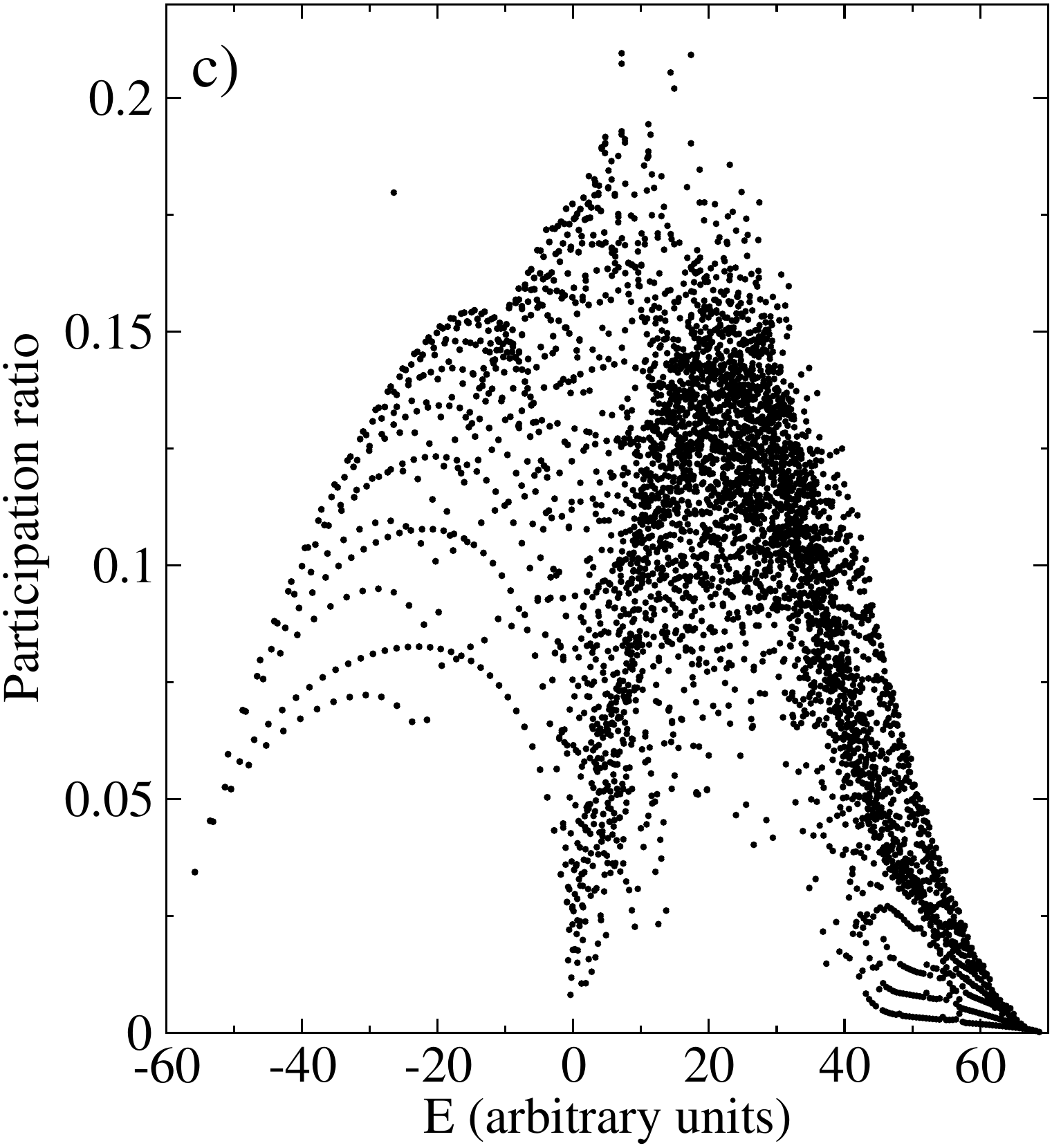}
\end{tabular}
\includegraphics[width=1\linewidth]{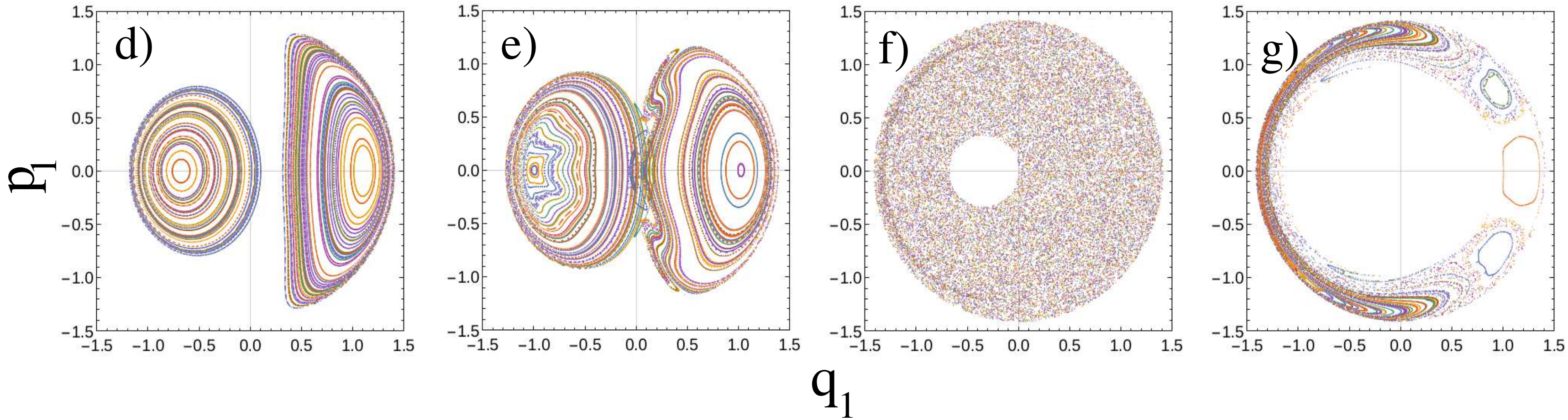}
\caption{Same as in Fig.~\ref{fig-comb-21} but for $y=0$, $y'=1$, and
  $x=1/2$. 
%XXXXXXXXXXXXXXXXXXXXXXXXXXXXXXXXXX  
Panels d), e), f), and g) are
  Poincar\'e sections for energies $E=-10$, $E=0$, $E=30$, and $E=50$.}  
\label{fig-comb-11}
\end{figure}

\begin{figure}
\begin{tabular}{ccc}
\includegraphics[width=0.30\linewidth]{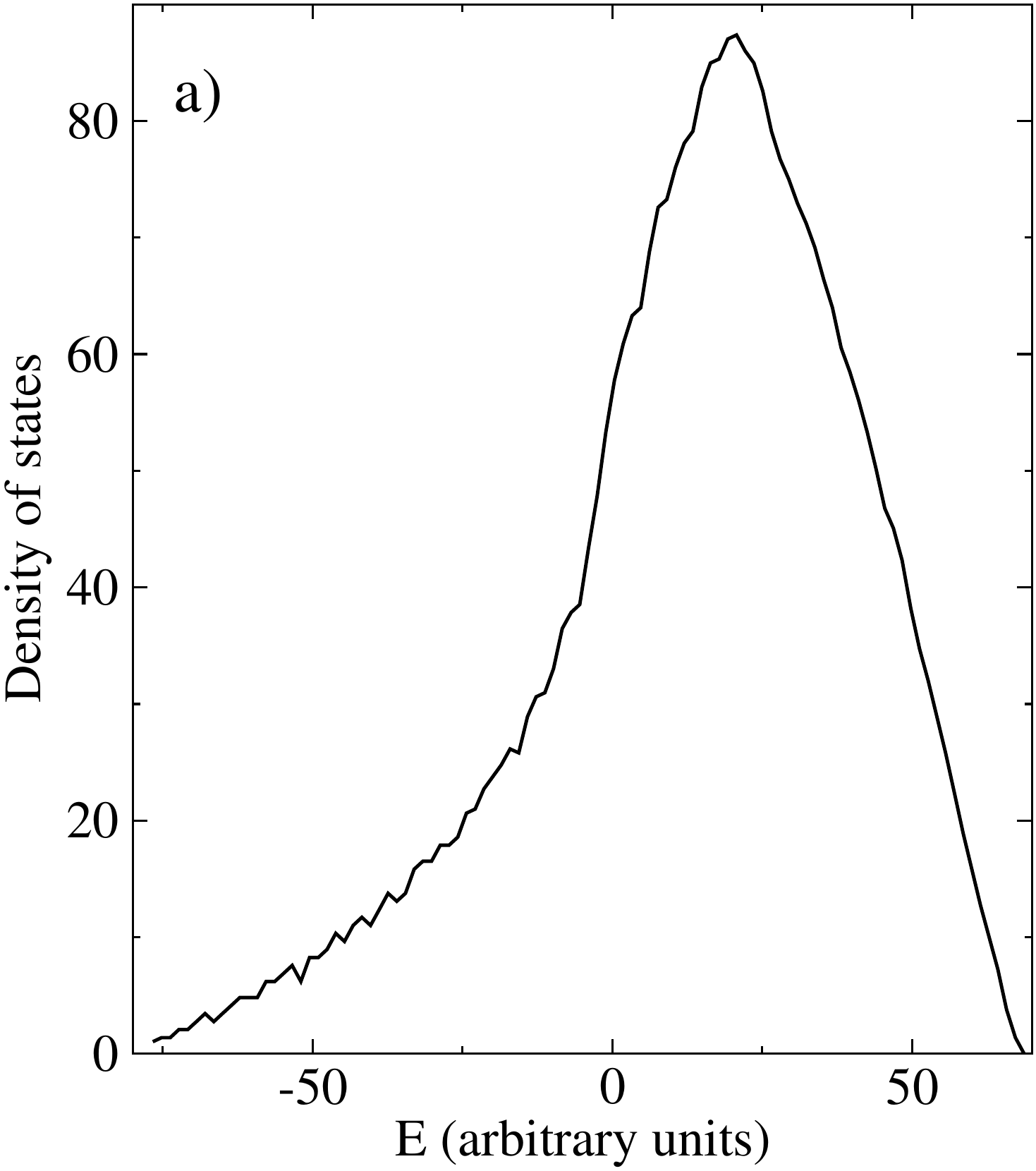}
&
\includegraphics[width=0.30\linewidth]{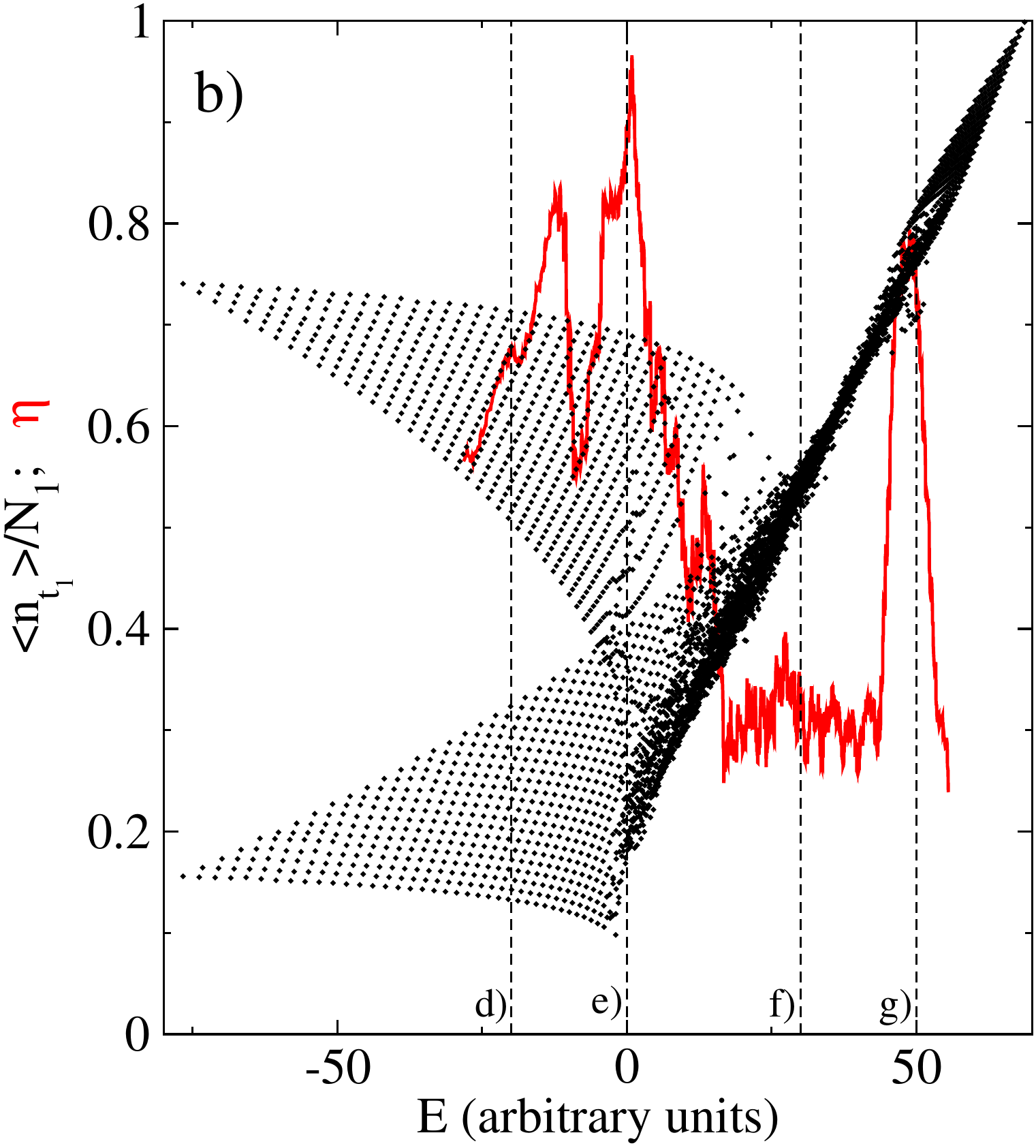}
&
\includegraphics[width=0.30\linewidth]{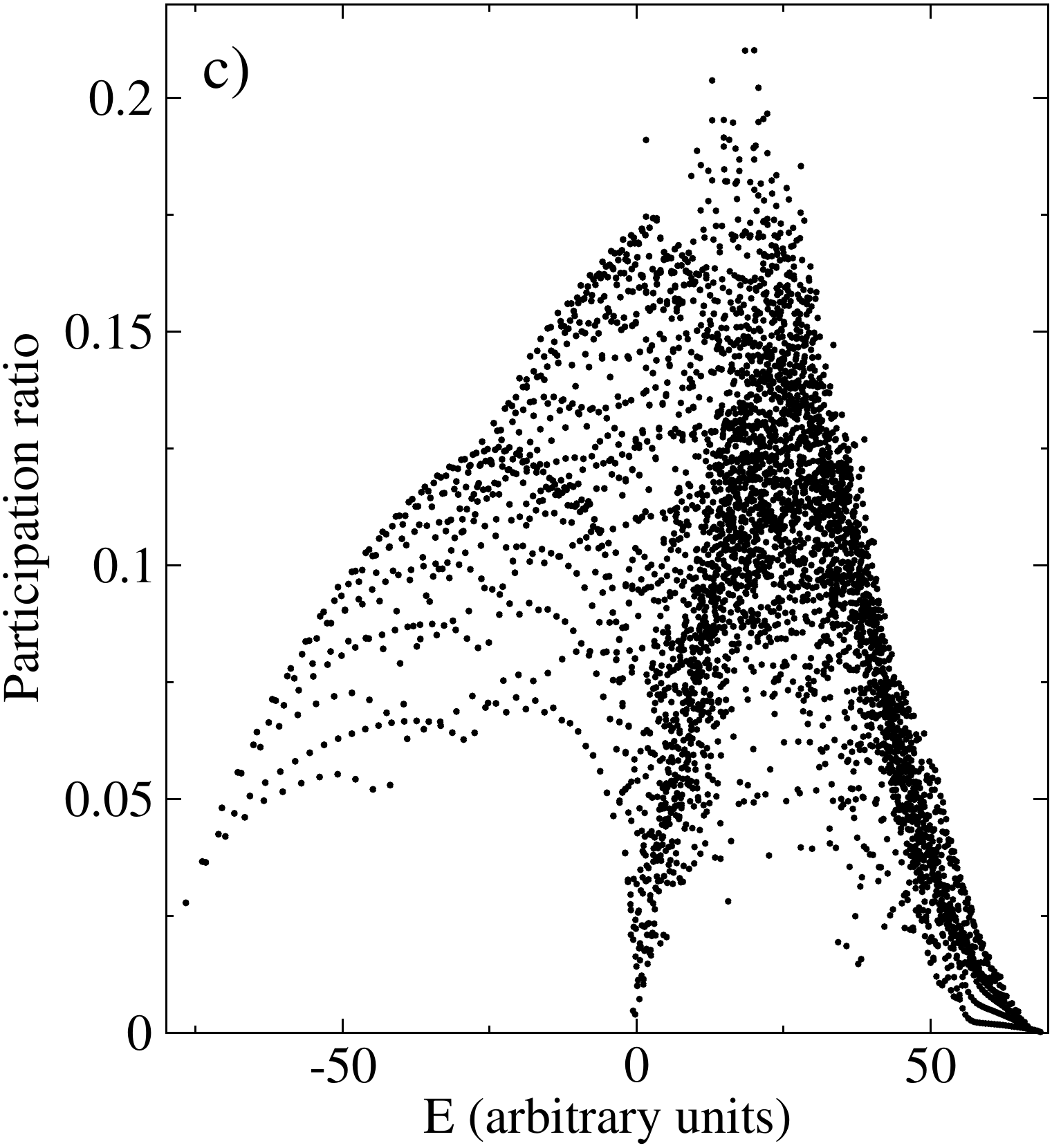}
\end{tabular}
\includegraphics[width=1\linewidth]{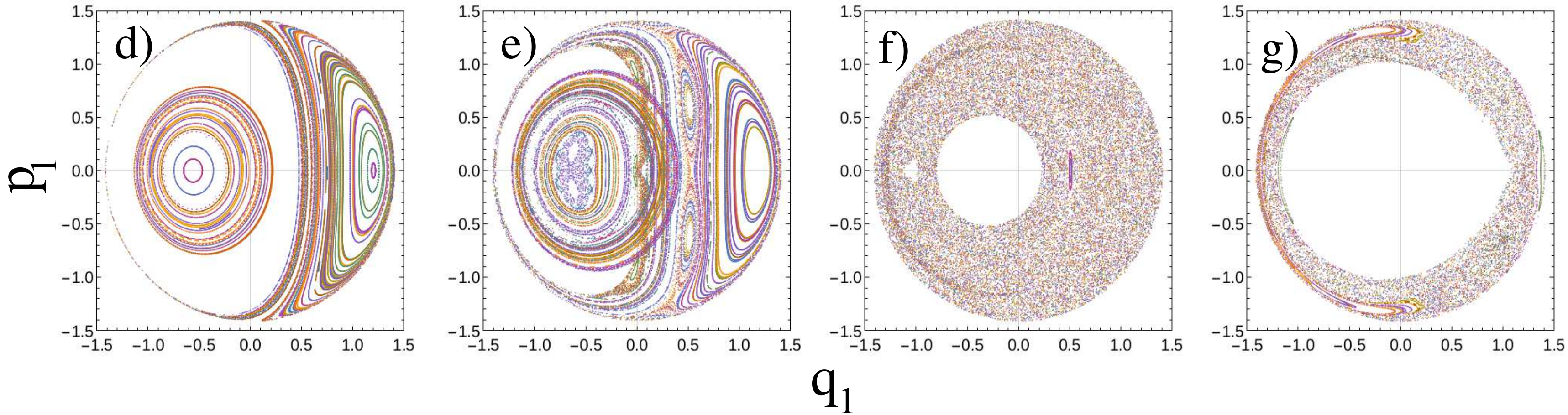}
\caption{Same as Fig.~\ref{fig-comb-21} but for $y=0$, $y'=3/2$, and
  $x=1/2$. 
%XXXXXXXXXXXXXXXXXXXXXXXXXXXXXXXXXX  
Panels d), e), f), and g) are
Poincar\'e sections for energies $E=-20$, $E=0$, $E=30$, and $E=50$.}  
\label{fig-comb-12}
\end{figure}

The three last cases correspond to Hamiltonians with $y=0$, {\it
  i.e.}, with $y_1=-y_2$, and, therefore, located on the gray vertical plane of the phase
diagram. As we proved in \cite{Garc16}, for values of 
$y'<1$ a second order QPT appears at $x_c=4/5$, for $y'=1$ the QPT shows a
divergence in $d^2E/dx^2$ also at  $x=4/5$, but for $y'>1$ the QPT
becomes of first order. Thus, next we try to disentangle whether or
not there is a different qualitative behaviour between these three
situations in terms of analyses of density of states, Peres lattice,
Poincar\'e section, and
participation ratio. In Fig.~\ref{fig-comb-13} the results for the case
of $y'=1/2$ are presented, in Fig.~\ref{fig-comb-11} the case of $y'=1$ is studied, and,
finally, in Fig.~\ref{fig-comb-12} the case of $y'=3/2$ is analyzed.   
%13
%\subsubsection{$y=0$, $y'=1/2$ ($y_1=1/2$, $y_2=-1/2$), $x=1/2$}
%\label{sec-13}

In Fig.~\ref{fig-comb-13} we consider the case $y=0$, $y'=1/2$ and
$x=1/2$ (point D in Fig.~\ref{fig-dia}). This corresponds to a
deformed ground state in a line that changing $x$ presents a second
order QPT at $x_c=4/5$. The energy surface has two deformed degenerate
minima (at $\beta$ and $-\beta$) separated by a spherical maximum. In
panel (a) the value of the density of states is presented as a
function of the excitation energy. A relatively broad peak is observed
with a maximum well above the energy of the ESQPT. In panel (b) the
Peres lattice for $\langle n_{t_1}\rangle/N_1 $ as a function of
excitation energy is presented. Note that because of the presence of
degenerated doublets, it is needed to include a tiny perturbation in the
Hamiltonian for slightly break the degeneracy and avoid random linear
combinations of states. Two branches of points are clearly
seen for energies below the ESQPT, one centered around $\langle
n_{t_1}\rangle/N_1\approx 0.3$ and the other around $\langle
n_{t_1}\rangle/N_1\approx 0.5$. These two branches correspond to the
two degenerated and symmetric minima for which $\langle n_{t_1} \rangle$
and $\langle n_{t_2} \rangle$ are interchanged. In both cases a clear
ordered pattern is observed. Note that for each energy there exist two
degenerated points, one in each branch. Above the ESQPT the order is
lost and the points are located in a more or less random way around a
straight line. As in previous cases, the ESQPT separates the regular
and the chaotic zones. In this sense, the ESQPT separates two
``shapes/phases'' of the system.  This is confirmed through the
calculation of the NNSD $\eta-$value, which shows a sudden decrease in
its value at zero energy, though with several oscillations, therefore
having a more regular behaviour below the ESQPT and more chaotic
above. However, even above the ESQPT, there is a more regular region
at $E\approx 50$ with $\langle n_{t_1}\rangle/N_1\approx 0.8$.
%XXXXXXXXXXXXXXXXXXXXXXXXXXXXXXXXXXXX
Panels
(d), (e), (f) and (g), which correspond to Poincar\'e sections with
energies $E=-10$, $E=0$, $E=20$, and $E=50$, respectively, confirm
the above findings: below the
ESQPT energy a regular behaviour exits (panel (d)), at the
ESQPT energy chaotic orbits star to appear (as shown in panel (e)),
and above the ESQPT energy, regions with chaotic (panel (f)) and partially regular
motion coexist (see panel (g)).
In panel (c) it is depicted the participation ratio  versus the
excitation energy. 
Below the ESQPT, the points (doubly degenerated) describe well
separated inverted parabolas with a minimum at zero energy. For
energies above the ESQPT, once more, the points define a very blurred
parabola.

%11
%\subsubsection{$y=0$, $y'=1$ ($y_1=1$, $y_2=-1$), $x=1/2$}
%\label{sec-11}

Next case to be analyzed is $y'=1$ which is probably the most exotic
one because the associated QPT at $x_c=4/5$ presents a divergence in 
$d^2E/dx^2$ (point E in Fig.~\ref{fig-dia}). However, as we will
explain, in fact, nothing special 
occurs in the spectrum. In Fig.~\ref{fig-comb-11}a we depict the
density of states and 
it looks like the one presented in Fig.~\ref{fig-comb-13}a. In
Fig.~\ref{fig-comb-11}b we present the Peres lattice and, once more,    
we have two branches for energies below the
ESQPT, one centered around 
$\langle n_{t_1}\rangle/N_1\approx 0.2$ and the other around $\langle
n_{t_1}\rangle/N_1\approx 0.6$. Note that for each energy there exist two
degenerated points. Above the ESQPT we obtain a cloud of points
scattered around a straight line. 
Here too, the NNSD $\eta-$value is plotted and once more it presents a
sudden lowering at the ESQPT energy, therefore pointing to a more regular
spectrum below the ESQPT, while more chaotic above. Note that to
calculate $\eta$ below the energy of the ESQPT, we only consider
states in one of the branches, otherwise the results are wrong. The decreasing occurs
in two steps, the first at zero energy and the second at $E\approx 20$. This fact is
a consequence of the coexistence of intruder and regular states in the
same energy region. In both
regions several oscillations are observed. Finally, there is an
striking increase of $\eta$ at $E\approx 50$, pointing to the presence
of a regular region above the ESQPT, as can be also observed in the
Peres lattice.
%XXXXXXXXXXXXXXXXXXXXXXXXXXXXXXXXXXXX
Once more, panels
(d), (e), (f) and (g), which correspond to Poincar\'e sections with
energies $E=-10$, $E=0$, $E=30$, and $E=50$, respectively, confirm
the above findings: below the
ESQPT energy a regular behaviour exits (panel (d)), at the
ESQPT energy chaotic orbits star to appear (as shown in panel (e)),
and above the ESQPT energy, regions with chaotic (panel (f)) and partially regular
motion coexist (see panel (g)).
In Fig.~\ref{fig-comb-11}c the
participation ratio is plotted, presenting well separated inverted
parabolas below the 
ESQPT, with the minimum at zero energy, while above the ESQPT a
single thick inverted parabola can be defined. 

%12
%\subsubsection{$y=0$, $y'=3/2$ ($y_1=3/2$, $y_2=-3/2$), $x=1/2$}
%\label{sec-12}

The latest case to be analyzed is $y=0$, $y'=3/2$ which is a line that
presents a first order QPT for $x_c\approx 4/5$. The value $x=1/2$
corresponds to an energy surface with two degenerate deformed minima
(point F in Fig.~\ref{fig-dia}), as in previous cases. The obtained figures,
\ref{fig-comb-12}a for the density of states, \ref{fig-comb-12}b for
the Peres lattice and, \ref{fig-comb-12}c for the participation ratio,
present only slightly differences with respect to
Fig.~\ref{fig-comb-11}. In particular, in the Peres lattice, the two
branches located below the ESQPT are centered around $\langle
n_{t_1}\rangle/N_1\approx 0.15$ and $\langle n_{t_1}\rangle/N_1\approx
0.75$, respectively. Here, the $\eta-$value changes rapidly when
crossing the ESQPT, separating the regular region below the ESQPT and
the chaotic one above. Once more, around $E\approx 50$ there is a
revival of the value of $\eta$ because of the occurrence of a more
regular region. The rest of features are qualitatively the same as in
Figs.~\ref{fig-comb-13} and \ref{fig-comb-11}, including participation
ratio and Poincar\'e sections.

\begin{figure}[hbt]
  \centering
\includegraphics[width=0.90\linewidth]{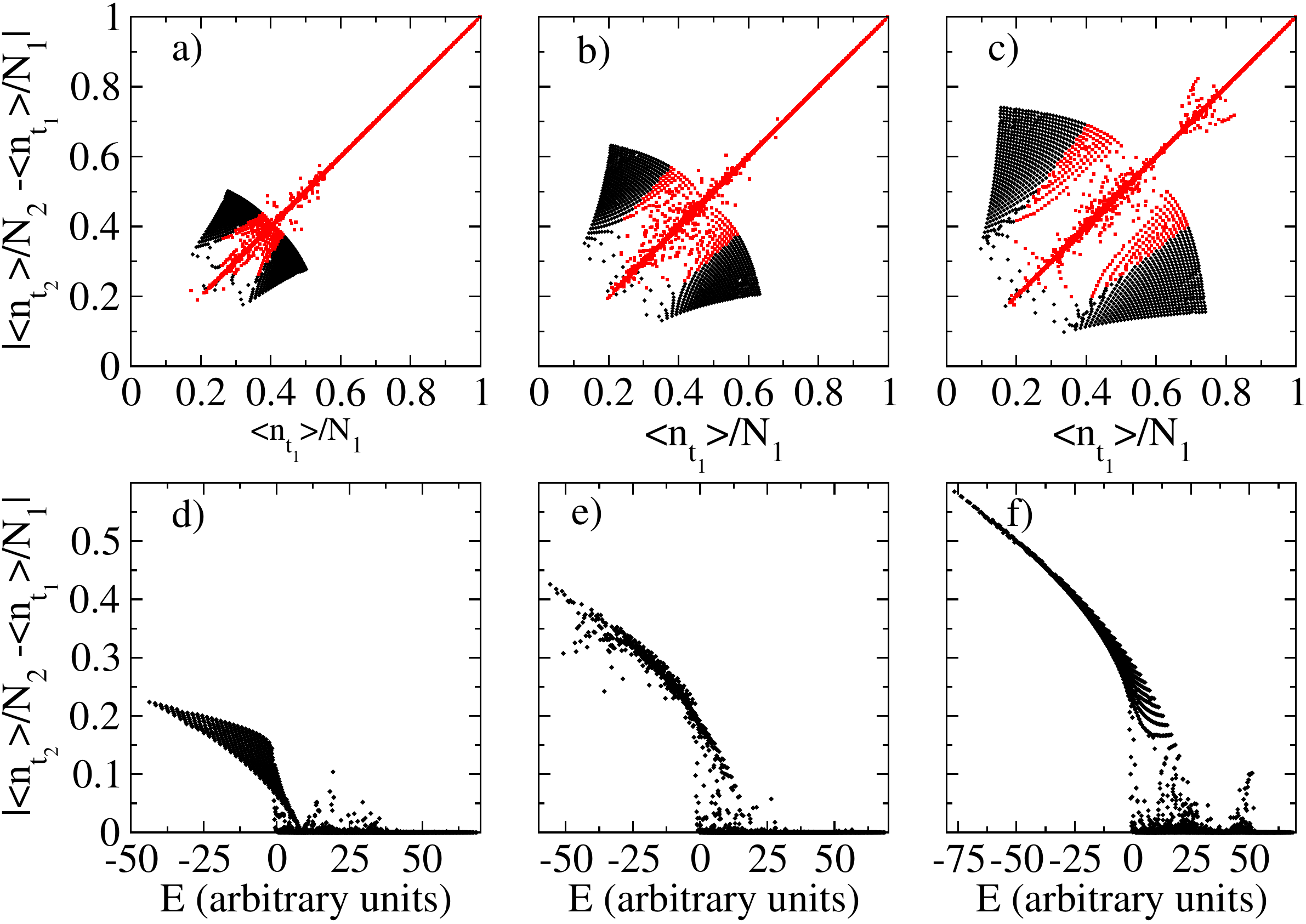}
\caption{$\langle n_{t_1} \rangle/N_1$ versus $\langle
  n_{t_2}\rangle/N_2$ for the Hamiltonians studied in
  Fig.~\ref{fig-comb-13} with parameters $x=1/2$, $y=0$, and $y'=1/2$
  (panel (a)), in
  Fig.~\ref{fig-comb-11} with parameters $x=1/2$, $y=0$, and $y'=1$
  (panel (b)), and in
  Fig.~\ref{fig-comb-12} with parameters $x=1/2$, $y=0$, and $y'=3/2$
  (panel (c)). Black points correspond to states with negative energy,
  while the red ones correspond to states with  positive energy.
  In panels  (d) , (e), and (f) we depict $\langle n_{t_2} \rangle/N_2$ - $\langle
  n_{t_1}\rangle/N_1$ as a function of the energy for the same
parameters than in panels (a), (b), and (c), respectively. } 
\label{fig-n1-n2}
\end{figure}

An alternative way of seeing the existence of different phases in the
spectrum is the use of a two-dimensional diagram representing $\langle
n_{t_1}\rangle/N_1$ versus
$\langle n_{t_2}\rangle/N_2$.  Moreover, we will see that $|\langle
n_{t_1}\rangle/N_1 - \langle n_{t_2}\rangle/N_2|$ can be considered as
an order parameter and therefore will be a useful tool to define the
phase of the system.   We will consider  the three last cases studied
with $y=0$. 
In  panels (a), (b), and  (c) of Fig.~\ref{fig-n1-n2} we depict $\langle
n_{t_1}\rangle/N_1$ versus 
$\langle n_{t_2}\rangle/N_2$, which corresponds to $y'=1/2$,  $y'=1$, 
and $y'=3/2$, respectively. The black points correspond to states with
negative 
energy, hence below the ESQPT, while the red ones correspond to states 
with  positive energy, therefore above the ESQPT. In this figure one
can easily single out the presence of two symmetric wells (they will
become asymmetric if $y\neq0$) that contain
the black points, while the rest of states (red points) are scattered
around the line $\langle n_{t_1} \rangle/N_1=\langle
n_{t_2}\rangle/N_2$. Note that $\langle n_{t_i} \rangle/N_i$ can be
connected with the order parameter through 
$\beta_i=\sqrt{\frac{\langle n_{t_i} \rangle/N_i}{1-\langle n_{t_i}
    \rangle/N_i}}$.  
% XXXXXXXXXXXXXXXXXXXXXXX
Panels (a), (b), and (c) of Fig.~\ref{fig-n1-n2}
provide a rough image of the 
available  position space for the states below or above the ESQPT. For
those below the ESQPT they are confined in two disjoint regions,  where
the potential in the Hamiltonian can be approximated by a quadratic form (one in each
well) that leads to a regular regime \cite{Emar,Stra15},
while above the
ESQPT, the harmonic approximation for the Hamiltonian is no
longer valid and a more chaotic behaviour is expected. 
%In this sense we can
%easily establish a qualitative connection between the presence of an ESQPT and
%the transition from a regular to a chaotic regime in the spectrum. 
In
panels (d), (e),  and (f) we plot $|\langle n_{t_2}\rangle /N_2 - \langle
n_{t_1}\rangle/N_1|$  as a function of the energy, corresponding to
$y'=1/2$,  $y'=1$, and $y'=3/2$, respectively. The variable  $|\langle
n_{t_2}\rangle/N_2 - \langle n_{t_1}\rangle/N_1|$ has the typical
behaviour of an order parameter, {\it i.e.}, it has a zero value in
the symmetric phase and different from zero in the non-symmetric one.
In our case, states with an energy below the ESQPT owns a finite
value of $|\langle n_{t_2}\rangle/N_2 - \langle
n_{t_1}\rangle/N_1|$  and it becomes zero for states with energy above
the ESQPT.

\section{Summary and conclusions}
\label{sec-conclusions}
In this work we have studied the onset of ESQPTs in a double Lipkin
Hamiltonian which resembles the consistent-Q Hamiltonian of the
interacting boson model. To find the presence of an ESQPT in the
spectrum we relied on the study of the density of states, the Peres
lattices, the Poincar\'e sections, the NNSD, and the participation ratio.

Taking into account the knowledge of the phase diagram of the model
\cite{Garc16}, we have selected particular points in the parameter
model space that correspond to non-symmetric (deformed) phases
and, therefore, should present an ESQPT in the spectrum at the energy
at which the potential energy surface has a maximum, that in our case
is always at zero energy. We have considered both cases with an ESQPT:
the one with just one deformed minimum and a spherical maximum, and
cases where two deformed minima appear separated  by a spherical maximum.
In this last case, a second family of states appears in the spectrum
when reaching the excitation energy of the second minimum.

Among the analyzed cases, first we started with Hamiltonians with
$u_{12}(2)$ as dynamical algebra and therefore with $j$ as good
quantum number. We have learnt that when looking into the spectrum as a
whole, one has to take into account that the symmetry of the
Hamiltonian is shaping the results. This is particularly evident for
the Peres lattice and the participation ratio. The conclusion for
these cases is that the position of the ESQPT is shifted to higher
energies as the value of $j$ decreases, but the main features of each
ESQPT are the same already described in the literature.  

Next we moved into cases where $u_{12}(2)$ symmetry was badly broken. Though
the lack of symmetry, the patterns that we observed were amazingly
similar to the previous cases and the zero energy clearly marks the
edge between two $shapes/phases$ with distinct patterns in the Peres
lattice and in the participation ratio. The change in the structure of
the Peres lattice, the Poincar\'e section, and in the NNSD ($\eta$) points to the passing from
a regular to a chaotic  or less regular regime once the ESQPT is crossed, although
above the ESQPT also appear regions with a strong regular
character. 
%XXXXXXXXXXXXXXXXXXXXXXXXXXX
The relationship between the appearance of an ESQPT and the onset of
chaos, seems to depend on the Hamiltonian parameters, and therefore
it is not possible to establish a clear connection between both
phenomena, as pointed out in Ref.~\cite{Chav16}. However, up to our
knowledge, all the cases studied in the literature, including ours, point towards a
regular behaviour for states below the ESQPT energy, {\it
  i.e.}~belonging to the non-symmetric phase, although the character
above the ESQPT energy strongly depends on the Hamiltonian.
Finally, the use of $|\langle n_{t_2}\rangle/N_2 - \langle
n_{t_1}\rangle/N_1|$ as order parameter clearly mark the presence of
an ESQPT at zero energy, which is deeply connected with the
change in the observed pattern in the diagram 
$(\langle n_{t_1}\rangle/N_1, \langle n_{t_2}\rangle/N_2)$ when
crossing the ESQPT energy.
  
In summary, in a compound system, as the two-fluid Lipkin model, the
existence of an ESQPT is self evident, though the value of the density
of states turns out not to be an appropriated quantity to single out its
presence. However, Peres lattices and participation ratio have
shown to be ideal tools to mark the presence of ESQPTs.

\section{Acknowledgment}
This work has been supported by the Spanish Ministerio de
Econom\'{\i}a y Competitividad and the European regional development
fund (FEDER) under Project No. FIS2014-53448-C2-1-P and FIS2014-53448-C2-2-P.

\end{document}